\documentclass[aps,floats,twocolumn,epsf,prb,showpacs]{revtex4}
\usepackage{graphicx}

\newcommand{\fig}[4][clip]{\begin{figure}[bt!]%
    \centerline{\includegraphics*[#1]{#2}}
    \caption{\sl\small\label{#3}#4}\end{figure}}
\newcommand{\gig}[4][clip]{\begin{figure}[t!]%
    \centerline{\includegraphics*[#1]{#2}}
    \caption{\sl\small\label{#3}#4}\end{figure}}
\newcommand{\tig}[4][clip]{\begin{figure}[t]%
    \centerline{\includegraphics*[#1]{#2}}
    \caption{\sl\small\label{#3}#4}\end{figure}}

\providecommand{\vev}[1]{\langle#1\rangle}

\begin{document}

\title{Dynamical mean field theory for manganites}
\author{Y.-F. Yang$^{a}$ and K. Held$^b$}
\affiliation{$^a$ Los Alamos National Laboratory, Los Alamos, New Mexico 87545, USA
\\
$^b$Institute of Solid State Physics, Vienna University of Technology, 1040 Vienna, Austria}
\date{Version 1, \today }

\begin{abstract}
Doped and undoped manganites are modeled by 
the coupling between itinerant $e_g$ electrons
and static $t_{2g}$ spins,   the Jahn-Teller and breathing phonon modes,
and the Coulomb interaction.
We   provide for a careful estimate of
all parameters and solve the
corresponding Hamiltonian by dynamical mean field theory.
Our  results for the one-electron
spectrum, the optical conductivity, the dynamic and static lattice
distortion, as well as  the Curie temperature show 
the importance of all of the above ingredients for a realistic
calculation as well as for describing
the unusual dynamical properties of manganites including
the insulating parent compound and the insulating-like
paramagnetic state of doped manganites. 
\end{abstract}

\pacs{71.27.+a, 71.10.Fd, 75.47.Gk}
\maketitle

\let\n=\nu \let\o =\omega \let\s=\sigma


\section{Introduction}
\label{Sec:Intro}

Manganites have attracted intensive interest during the last decades due to their extraordinary properties including the colossal magnetoresistance (CMR) 
\cite{Helmolt1993,Chahara1993,Jin1994}. These materials have the chemical composition T$_{1-x}$D$_x$MnO$_3$, where T is a trivalent rare earth ion (T = La, Pr, Nd, $\dots$) and D is a divalent alkali ion (D = Ca, Sr, $\dots$) and crystallize in a cubic perovskite structure with a possible distortion at low temperatures, see 
Fig.\ \ref{Fig:sec:intro:structure}, which albeit can be distorted at low temperatures. Soon after CMR was discovered, various phase diagrams as a function of temperature and magnetic field or doping concentration $x$ were established \cite{Tokura1994,Urushibara1995,Kuwahara1995,Tomioka1995a,Tomioka1995b,
Tomioka1996,Tomioka1997,Moritomo1997,Mukhin1998}; also see the review articles Refs.\ \onlinecite{Tokura1996,Tokura1999,Salamon2001,Dagotto2001}. For the parent compound LaMnO$_3$, an insulator-to-metal transition was found upon applying pressure \cite{Loa2001}; and a charge/orbital-ordered phase has been reported in a large number of perovskite (T$_{1-x}$D$_x$MnO$_3$) and layered (T$_{1-x}$D$_{1+x}$MnO$_4$) manganites, depending on the effective
bandwidth and the quenched disorder \cite{Tomioka2004,Mathieu2006}. 
The dynamical properties of the paramagnetic insulating are very unusual
as is reflected  in a spectral function $A(\omega)$ with a very low spectral weight at the Fermi level $E_F$ irrespectively of $x$, as indicated by photoemission and X-ray absorption experiments \cite{Bocquet1992,Chainani1993,Saitoh1997,Park1996}. Similarly, the optical conductivity $\sigma(\omega)$ shows a very low spectral weight up to an energy scale of $\sim\! 1\,$eV \cite{Okimoto1995,Quijada1998,Jung1998,Takenaka1999}. Besides, the ferromagnetic metallic phase is an atypical (bad) metal \cite{Okimoto1995}.

\tig[clip=true,width=5.2cm,angle=0]{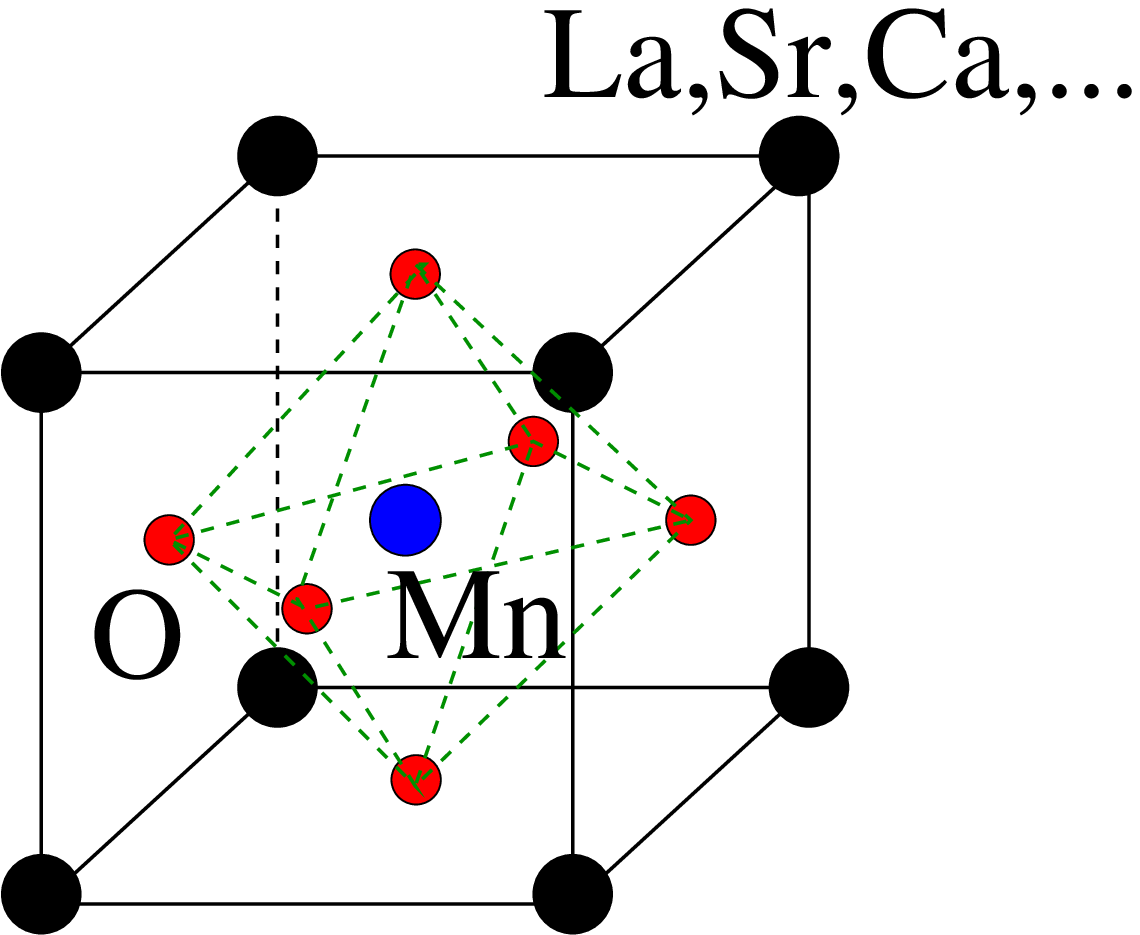}
{Fig:sec:intro:structure}
{Sketch of the cubic perovskite unit cell for manganites. The dashed lines
indicate the basic MnO$_6$ octahedron.}

A physical understanding of these properties is difficult due to the internal complexity resulting from the interplay between charge, spin, orbital and lattice degrees of freedom \cite{Millis1998,Tokura2003}. A basis ingredient of a  theoretical description is the separation of the electrons within the five $d$ orbitals into a localized $t_{2g}$ spin of length  $|{\bf S}|=3/2$ and $n=1-x$ itinerant $e_g$ electrons, coupled to the $t_{2g}$ spin by Hund's exchange. This  the basis of the so-called "double exchange" \cite{Zener1951a,Zener1951b}
which led to the (ferromagnetic) Kondo lattice model \cite{Kubo1972} and explains ferromagnetism in doped manganites and the charge-ordered phase at x=0.5 \cite{Goodenough1955}. A spin-canted state was suggested later \cite{Gennes1960}. However, this double exchange modeling disagrees with the
experiment in many aspects including the CMR. Hence, the importance of the
Jahn-Teller phonon modes and their couling to the $e_g$ electrons was stressed \cite{Millis1995} and studied 
\cite{Roder1996,Millis1996a,Millis1996b,Millis1996c}. However, while it can describe  insulating-like behavior for large electron-phonon coupling when electrons are trapped as lattice polarons \cite{Millis1996a}, the Kondo lattice model extended by Jahn-Teller phonons still fails to produce a large magnetoresistance at finite doping. Another important ingredient for the physics of manganites is the local Coulomb interaction between the $e_g$ electrons as was pointed out in \cite{Held2000} and in realisitic {\em ab initio} calculations \cite{Banach2004,Zenia2005,Yamasaki2006,Yin2006}. Of these {\em ab initio} calculations, the combination of the local density approximation (LDA) and dynamical mean field theory (DMFT) \cite{Anisimov1997,Lichtenstein1998,Katsnelson2000,Held2003,Kotliar2006,Held2007} has the least tendency to overestimate the formation of an insulating state since the DMFT electron dynamics also avoids the cost of double occupations in the paramagnetic metallic state. The corresponding LDA+DMT results for manganites \cite{Yamasaki2006} show that only the  combined localization effect
of Coulomb interaction and (static) Jahn-Teller distortion makes the undoped parent compound an insulator, and leads to a correct desription of the pressure induced insulator-to-metal transition \cite{Loa2001}. Similarly, for the doped compound, both mechanisms work together in localizing the $e_g$ electrons giving rise to a proper description of the unusual dynamical properties of doped
manganites and the large CMR in a wide range of doping \cite{Yang2006}.

In this paper, we present a detailed theoretical investigation of manganites using DMFT \cite{Metzner1989,Georges1992b,Georges1996} to solve a realistic model of manganites which supplements the Kondo-lattice model by Jahn-Teller and breathing mode phonons as well as by the local Coulomb interaction between the $e_g$ electrons; first results have been presented in \cite{Yang2006} also see the recent DMFT calculation \onlinecite{Lin2008}. The paper is organized as following: Section \ref{Sec:Model} introduces the Hamiltonian
and discusses how realistic parameters are choosen. Section \ref{Sec:Parent} is devoted to the undoped parent compound LaMnO$_3$ focusing in Section \ref{Sec:Parent:PM} on how the Jahn-Teller phonons, for a cubic lattice, give rise to a very similar spectrum as that of  LDA+DMFT for the experimental static Jahn-Teller distortion. This is supplemented by Section \ref{Sec:Parent:PT}
studying the symmetry breaking towards a static Jahn-Teller distortion. Section \ref{Sec:doped} presents results for doped mangaites including a detailed study of the paramgnetic insulating-like phase in Section \ref{Sec:doped:PI} and of the phase transition towards ferromagnetism in Section \ref{Sec:doped:FM}. Section \ref{Sec:Breathing} includes the effect of the breathing mode with
an estimate of the Coulomb repulsion  and the electron-phonon coupling strength in Section  \ref{Sec:Breathing:Parameters} and the corresponding optical conductivity in Section \ref{Sec:Breathing:Optics}. Finally, Section \ref{Sec:Conclusion} gives a summary of the results obtained.

\section{Realistic model and parameters}
\label{Sec:Model}
For the realistic microscopic modeling of manganites, we employ the Hamiltonian
\begin{eqnarray}
H=&-&\sum_{\vev{\vev{ij}};\mu\nu\sigma}t^{ij}_{\mu\nu}
c^{\dagger}_{i\mu\sigma}c_{j\nu\sigma} 
-2{\cal J} \sum_{i;\mu} {\bf s}_{i\mu}\cdot{\bf S}_{i}\nonumber \\
&+& U\sum_{i;\mu}n_{i\mu\uparrow}n_{i\mu\downarrow}
+\sum_{i;\sigma\tilde{\sigma}}(U'-\delta_{\sigma\tilde{\sigma}}J)
n_{i1\sigma}n_{i2\tilde{\sigma}}
\nonumber \\
&+&\sum_{i;a}\left(\frac{P_{ai}^2}{2}+\frac{\Omega^2}{2}Q_{ai}^2\right).
\nonumber \\
&-& g\sum_{i;\mu\nu\sigma}c^{\dagger}_{i\mu\sigma}
(Q_{1i}\mathcal{I}+Q_{2i}\tau^x+Q_{3i}\tau^z)_{\mu\nu}c_{i\nu\sigma}.
\label{Eq:sec:doped:dmft:parameters:Ham}
\end{eqnarray} 
Here, $c^{\dagger}_{i\mu\sigma}$ and $c_{i\mu\sigma}$ are the fermionic
creation and annihilation operators for electrons at site $i$ within
$e_g$-orbital $\mu$ and with spin $\sigma$; ${\bf s}_{i\mu}$ is the
corresponding spin operator 
${\bf s}_{i\mu}=\sum_{\sigma_1\sigma_2}c^{\dagger}_{i\mu\sigma_1}
\frac{{\mathbf \tau}_{\sigma_1\sigma_2}}{2}c_{i\mu\sigma_2}$
with Pauli matrices ${\mathbf \tau}$.

The first line of Hamiltonian (\ref{Eq:sec:doped:dmft:parameters:Ham}) forms the ferromagnetic Kondo lattice model with a coupling ${\cal J}$ of the $e_g$ spin to the localized $t_{2g}$ spin ${\bf S}_{i}$. For the cubic lattice, the hopping elements  $t^{ij}_{\mu\nu}$ can be, to a good approximation,  restricted to neighboring sites $i$ and $j$ in the $x$, $y$ and $z$ direction with an orbital matrix for the two ($e_g$) orbitals $d_{x^2-y^2}$ and $d_{3z^2-r^2}$ indexed by $\mu$ (and $\nu$) 
\begin{equation}
t^x=t_0\left(\begin{array}{cc}
\frac34 & -\frac{\sqrt{3}}{4} \\
-\frac{\sqrt{3}}{4} & \frac14
\end{array}\right),
\;\;\;\;
t^y=t_0\left(\begin{array}{cc}
\frac34 & \frac{\sqrt{3}}{4} \\
\frac{\sqrt{3}}{4} & \frac14
\end{array}\right),\nonumber
\end{equation}
\begin{equation}
t^z=t_0\left(\begin{array}{cc}
0 & 0 \\
0 & 1
\end{array}\right).
\label{Eq:sec:model:tight}
\end{equation}
The constant  $t_0=W/6=0.6\,$eV was estimated from LDA which gives a bandwidth $W= 3.6\,$eV for cubic LaMnO$_3$ \cite{Yamasaki2006}. From the 2.7 eV splitting 
between spin-up and spin-down $e_g$ bands in the corresponding ferromagnetic LDA calculatation on the other hand, the value of coupling to the $t_{2g}$ spins
was estimated $2{\cal J}|{\bf S}_{i}|=2.7\,$eV.

The second line of Hamiltonian (\ref{Eq:sec:doped:dmft:parameters:Ham})
describes the Coulomb repusion of two $e_g$ electrons on the same lattice site $i$, consisting of the intra-orbital Coulomb repulsion $U$, the inter-orbital repulsion $U'$ and the Hund's rule energy gain $J$ for two parallel $e_g$ spins. The Coulomb interaction can be estimated from the experimental spectrum, combining photoemission spectroscopy (PES) and x-ray absorption spectroscopy (XAS) \cite{Park1996}. Accounting for the crystal field splitting between $e_g$ states and $t_{2g}$ states of $1-2\,$eV, we obtain an average Coulomb interaction $\bar{U}=3-4\,$eV, which is also supported by spectral ellipsometry for LaMnO$_3$ \cite{Kovaleva2004}. Since $U'=U-2J$ by symmetry, we have the relation 
\begin{equation}
\bar{U}=\frac{U+U-2J+U-3J}{3}=U-\frac{5}{3}J. 
 \end{equation}
Taking a value $J=0.75\,$eV, which is slightly smaller than the value for the five band model obtained by constrained LDA \cite{Satpathy1996a} and which agrees with our own estimate for ${\cal J}$, all Coulomb interactions are determined. Unless noted otherwise, we employ $U'=3.5\,$eV.

The third line is the Hamiltonian for the three most important phonon modes $Q_a$, i.e., the breathing mode $Q_1$ and the two Jahn-Teller modes  $Q_{2,3}$ as illustrated in Fig.\ \ref{Fig:sec:model:PE}. The only free parameter of this part of the Hamiltonian (\ref{Eq:sec:doped:dmft:parameters:Ham}) are the three phonon frequencies which, following  Ref.\ \onlinecite{Iliev1998}, can be obtained from the Raman spectrum for LaMnO$_3$. The Raman spectrum gives very similar values
for the Jahn-Teller mode frequency $\Omega_{JT}=0.07\,$eV and that of the breathing mode  $\Omega_{br}=0.08\,$eV so that we take a unique value $\Omega=0.07\,$eV in this paper.

\tig[clip=true,width=8.5cm,angle=0]{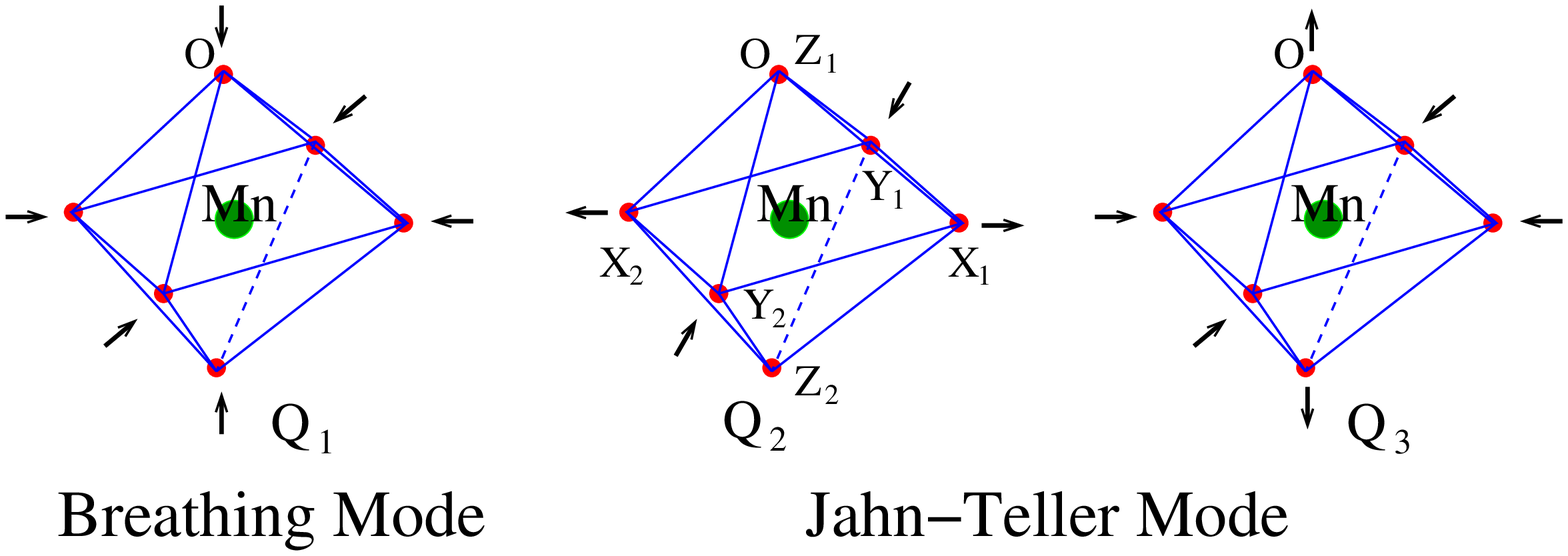}
{Fig:sec:model:PE}
{Vibration of the oxygen octahedra around the manganese ion for the breathing mode $Q_1$ and the two Jahn-Teller modes $Q_2$ and $Q_3$.}

Finally, the fourth line is the electron-phonon coupling with the breathing mode coupling to the electron density and the two Jahn-Teller modes coupling to the difference in $e_g$ occuption. Avoiding to use too many parameters, we take, as for the frequency, a unique coupling strength $g$. For the Jahn-Teller modes it
is related to the static Jahn-Teller energy $E_{JT}=g^2/2\Omega^2$, the ground state energy if only the Jahn-Teller coupling and phonon energy is present. One can try to determine the Jahn-Teller coupling stength $g$ from the lattice distortion. A distortion  of 0.1 \AA{} which is consistent with some LDA calculations \cite{Pickett1996,Yin2006} leads to $E_{JT}=0.25\,\text{eV}$ and hence $g=0.05\,$eV$^{3/2}$ for $\Omega=0.07\,$eV. However, recent x-ray powder diffraction and neutron powder diffraction measurements found a much larger distortion \cite{Chatterji2003}, which would result in unrealistically large values for $g$ or frequencies  $\Omega$ at odds with the Raman frequencies.
We hence attribute these larger distortions to cooperative lattice effects and the quadratic vibronic coupling to the electronic degrees of freedom. Possibly also enhancement effects due to electronic correlations play a role. As a consequency the precise value of $g$ is an open issue and we have hence done calculations for various values of $g$. In  section \ref{Sec:Breathing:Parameters}, we provide for a new estimate of the two parameters with the biggest uncertainty, i.e., $U$ and $g$ on the basis of ther experimental gap of undoped LaMnO$_3$ and the resistitvity for doped LaMnO$_3$.

For the folowing results, Hamiltonian (\ref{Eq:sec:doped:dmft:parameters:Ham}) is solved using DMFT \cite{Metzner1989,Georges1992b,Georges1996} with Hirsch-Fye \cite{Hirsch1986} Quantum Monte Carlo (QMC) simulations supplemented by the Blankenbecler-Scalapino-Sugar algorithm \cite{Blankenbecler1981} for Holstein phonons. To this end, the inverse temperature $\beta$ is discretized into $L$ time slices $\tau_l=(l-1)\Delta\tau,\,\,(l=1,\cdots,L+1)$ of size $\Delta\tau=0.25\,$eV$^{-1}$ and the $t_{2g}$ spin is assumed to be classical. Let us briefly discuss some aspects of the phonon fields since these are less commonly simulated in DMFT. The phonon field can be described by a classical field $\phi_l$ with  boundary condition $\phi_{L+1}=\phi_1$ \cite{Blankenbecler1981}. In the effective action, the kinetic and potential
energy of the phonons [third line in  (\ref{Eq:sec:doped:dmft:parameters:Ham})]
hence become 
\begin{eqnarray}
K(\{\phi_l\})&=&\frac{\Delta\tau}{2}\sum_{l=1}^L\left[\left(\frac{\phi_{l+1}-\phi_{l}}{\Delta\tau}\right)^2+\omega^2\phi_l^2\right].
\label{Eq:DMFT:DMFT:holstein}
\end{eqnarray}
We employ local  updates for one time slice $l$ and global updates for all time slices, both having the form $\phi_l^\prime=\phi_l + (x-1/2) \delta\phi$ with a random number $x\in[0,1)$ and a properly choosen amplitude $\delta\phi$  different for local and global updats. For low temperatures $T$ and large electron-phonon couplings $g$, the probability to go through the QMC sampling 
from  a region with large lattice distortion (occupied with one $e_g$ electron)
to a region with small lattice distortion without $e_g$ electron becomes very low. This is quite similar to the Hubbard model where for large $U$ and small $T$, the tranisition between configurations with predominately spin-up and spin-down becomes very rare. Fortunately, the physics in both situations is rather simple and a proper averaging between spin-up and spin-down respectively
small and large lattice distortion is easily possible. In the latter (manganites) case, the probability ${\cal P}=x$ for unoccupied, undistorted  sites is physical meaningful and also obtained by an extrapolation from higher temperatures.

From the DMFT one-electron spectral function the optical conductivity is calculated from the simple bubble diagram given by two Green functions multiplied by two group velocities (derivates of the dispersion relation). For the magnetic transition temperature $T_c$, the two particle Green function
and from that the inverse susceptibiliy crossing zero at $T_c$ were calculated.

\section{Insulating parent compound}
\label{Sec:Parent}
\subsection{Paramagnetic phase}
\label{Sec:Parent:PM}
We start our study with the undoped parent compound LaMnO$_3$, which for a static Jahn-Teller distortion was studied in Ref.\ \cite{Yamasaki2006} using LDA+DMFT. Fig.\ \ref{Fig:sec:undoped:DMFT:spectral} shows the spin- and orbital-averaged spectral density
$ A(\omega)=-\frac{1}{4\pi}\sum_{\mu\sigma}\Im G^{\mu\sigma}(\omega)$
in the paramagnetic phase with the Green functions $G^{\mu\sigma}(\omega)$
obtained for real frequencies $\omega$ by the maximum entropy method 
\cite{Jarrell1996}.
\tig[clip=true,width=5.5cm,angle=270]{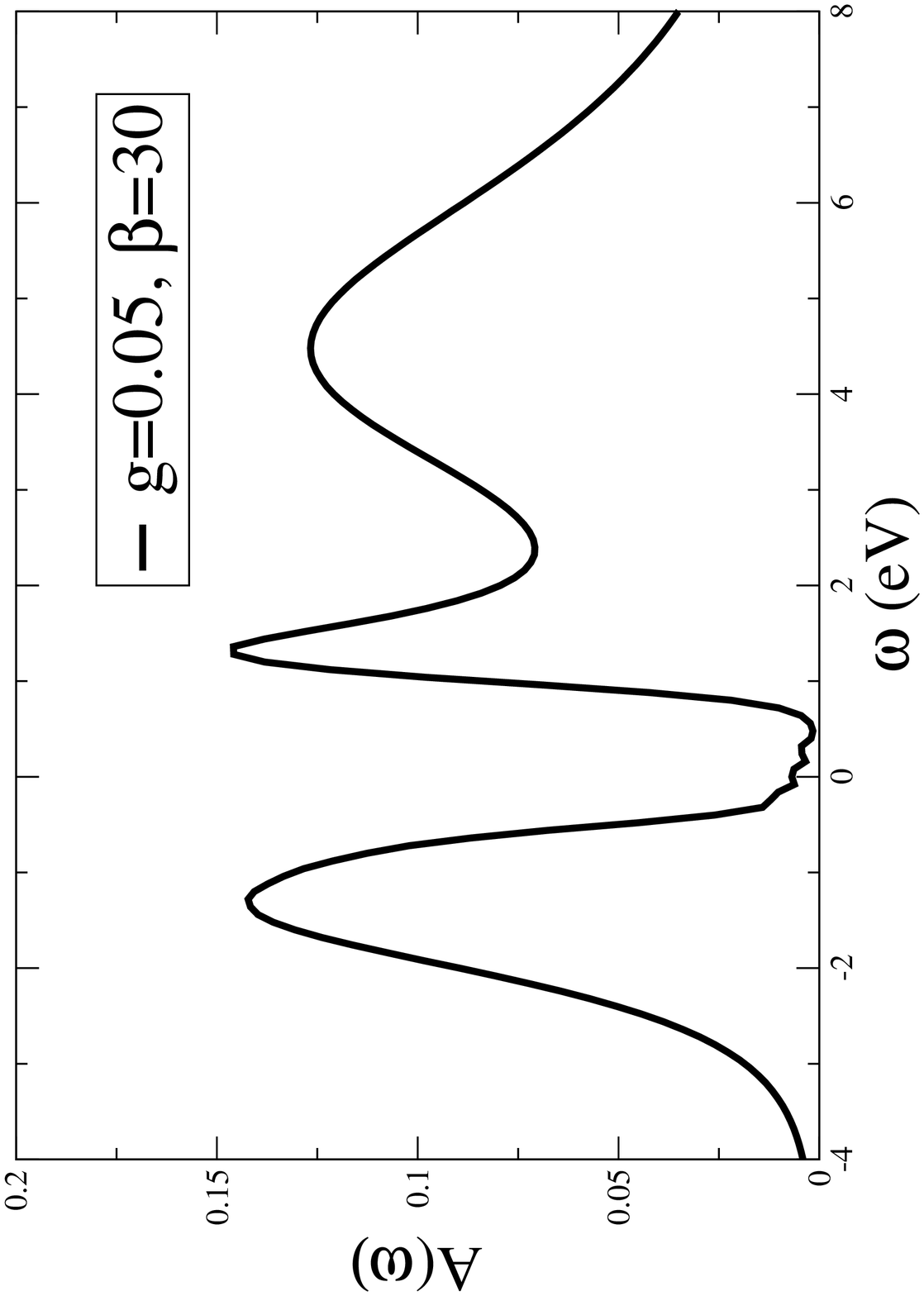}
{Fig:sec:undoped:DMFT:spectral}
{Local spectral density in the paramagnetic phase of LaMnO$_3$ calculated for
$g=0.05\,$eV$^{3/2}$, $U'=3.5\,$eV and $\beta=30\,$eV$^{-1}$. All other parameters are fixed as discussed in Section \ref{Sec:Model}. The breathing mode phonon will only be included later in Sec.\  \ref{Sec:Breathing}.} 
It shows a three preak structure very similar to the previous LDA+DMFT results \cite{Yamasaki2006} even though no static Jahn-Teller distortion is present. The reason for this is that the lattice dynamics leads to a similar (averaged) lattice distortion as in the static case, see Fig.\ \ref{Fig:sec:undoped:DMFT:distribution}. At a given time shot (or for a given QMC configuration), the two $e_g$ levels are Jahn-Teller split with the precise decomposition of the lower (and higher) lying orbital depending on the sign 
and admixture of the $Q_2$ and $Q_3$ Jahn-Teller modes. The first peak around -1.3 eV corresponds to electronic states occupying the lower of the two $e_g$ levels, and the second one around +1.3eV to adding an electron in the higher lying orbital. The third peak around +4.5 eV corrsponds to excitations
to states with antiparallel  $t_{2g}$ spin, whereas the first two peaks have parallel $t_{2g}$ spin. In Fig.\ \ref{Fig:sec:undoped:DMFT:spectral}, there is still some spectral weight in the gap which may stem from the phonon sideband \cite{Edwards2002}. The energy gap can hence only be obtained approximately.  It is comparable to the  experimental gap of about 1 eV \cite{Jung1998,Takenaka1999}. The consistency between the previous LDA+DMFT and the present DMFT model calculations supports our realistic microscopic model (\ref{Eq:sec:doped:dmft:parameters:Ham}) for describing the electronic behavior of LaMnO$_3$ and demonstrates once again that the insulating ground state in LaMnO$_3$ at ambient conditions results from the combination of the Coulomb interaction and the Jahn-Teller coupling in addition to the Hund's coupling between the $e_g$ and $t_{2g}$ spins.

Fig.\ \ref{Fig:sec:undoped:DMFT:distribution} plots the probability distribution $P(Q)$ for a given lattice distortion in the range $[Q,Q+dQ]$. Note that the lattice distortion are converted into units of \AA{} by multiplying a factor of $\hbar/\sqrt{M}$ where $M$ is the mass of oxygen atom and that for a finite $\Delta \tau$ the two modes are not exactly symmetric. 
\fig[clip=true,width=5.5cm,angle=270]{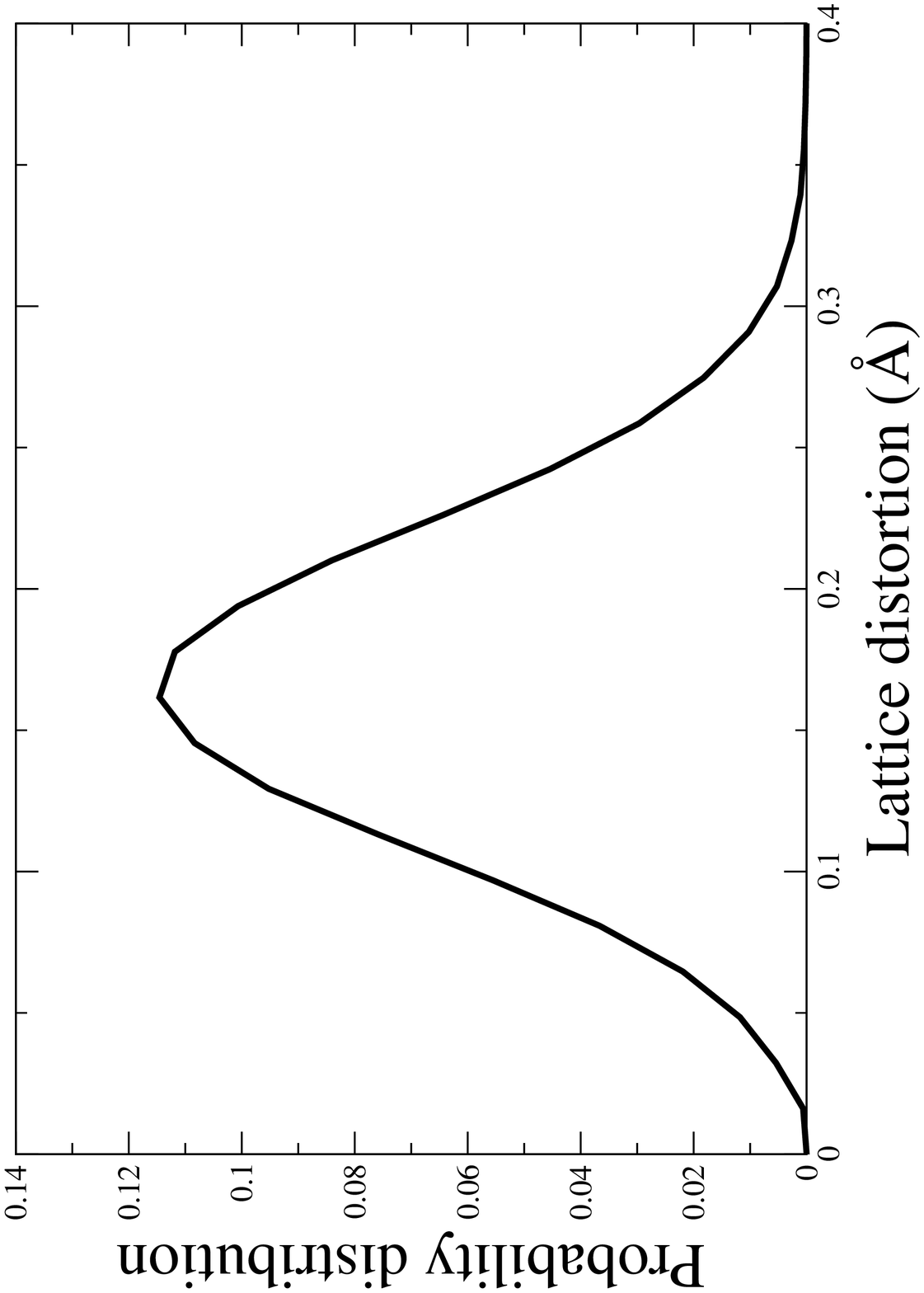}
{Fig:sec:undoped:DMFT:distribution}
{Probability distribution (in arbitrary unit) of the lattice distortion for the Jahn-Teller coupling $g=0.05\,$eV$^{3/2}$, Coulomb repulsion $U'=3.5\,$eV and
inverse temperature $\beta=16\,$eV$^{-1}$.}

We see that $P(Q)$ has one broad peak located at around 
\begin{equation}
\bar{Q}=\frac{\int\,dQ Q P(Q)}{\int\,dQ P(Q)}=0.167\,\AA,
\end{equation}
which corresponds roughly to $Q^\star=g/\Omega^2\approx 0.15\,$\AA{} obtained
for a single-site model. As we have discussed, the discrepancy from the neutron experiment \cite{Chatterji2003} is probably due to the higher order quadratic vibronic coupling and the cooperative effect between the adjacent MnO$_6$ Octahedra that share a common oxygen atom \cite{Popovic2000}. A recent LDA+U calculation has taken into account these effects and produced the correct experimental results \cite{Yin2006}. Cluster extensions are required for further realistic DMFT calculations.

\subsection{Structural transition}
\label{sec:undoped:DMFT:transition}
\label{Sec:Parent:PT}

Experimentally a structural first-order phase transition is observd  at
$T_{OO}\approx 740\,$K with an abrupt volume contraction \cite{Chatterji2003}. The lattice is nearly cubic above $T_{OO}$ but has a strongly distorted orthorhombic structure due to the static Jahn-Teller distortion below $T_{OO}$. The structural transition is accompanied by an orbital order-disorder transition. The low temperature phase shows a staggered ordering of the $d_{3x^2-r^2}$ and $d_{3y^2-r^2}$ orbitals in the $a$-$b$ plane which repeats itself along the $c$-direction.

With a slight modification of the single-impurity DMFT (QMC) algorithm, we can
study a hypothetical antiferromagnetic orbital ordering on an $AB$ lattice and
draw some conclusions about the structural transition in LaMnO$_3$. To study the alternating  orbital ordering, the cubic lattice is separated into two sublattices A and B with oposite behavior of the two $e_g$ orbitals:
$\Sigma^A_{1(2)}(\omega)=\Sigma^B_{2(1)}(\omega)$. Due to the symmetry, we only need to take care of a single lattice site of either type in DMFT.

Fig.\ \ref{Fig:sec:undoped:DMFT:spectralAB} shows the local spectral densities
of two $e_g$ orbitals for $U'=3.5\,$eV, $T=0.05\,$eV and $g=0.05\,$eV$^{3/2}$.
At this temperature, the orbital symmetry is strongly broken: One orbital is occupied with almost one electron, while the other orbital is only slightly occupied. Fig.\ \ref{Fig:sec:undoped:DMFT:spectralAB} also indicates the contributions from different spin and orbital components. A direct consequence of the orbital order is, due to the Jahn-Teller coupling, a corresponding lattice distortion. Averaged over the two orbital, the spectrum is actually very similar to that of Fig. \ref{Fig:sec:undoped:DMFT:spectral} with the two peaks at $-1\,$eV and $2\,$eV stemming from the spin states parallel to the local $t_{2g}$ spins. These two peaks are split by a combination of Jahn-Teller coupling and Coulomb interaction. 

\tig[clip=true,width=5.5cm,angle=270]{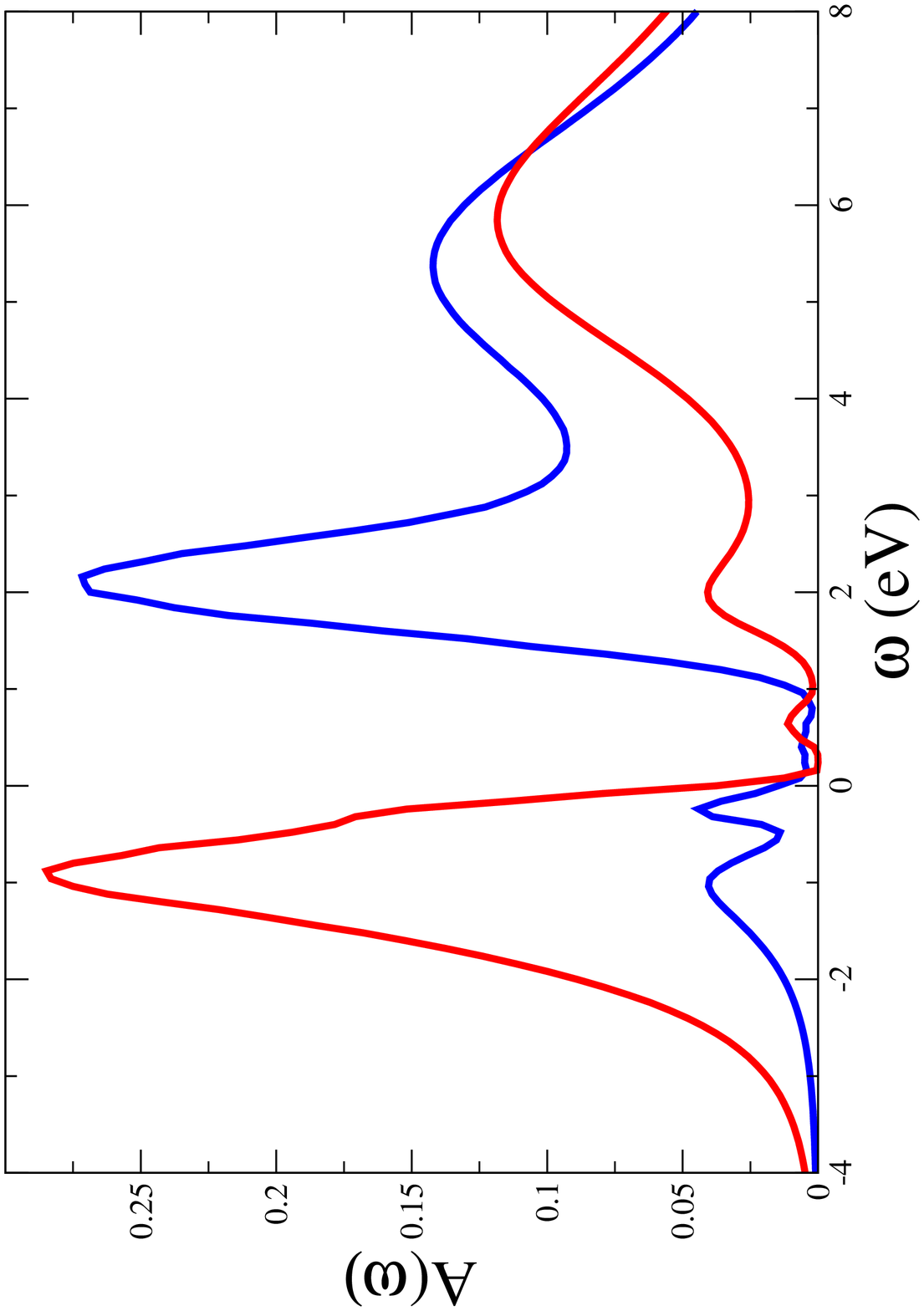}
{Fig:sec:undoped:DMFT:spectralAB}
{Local spectral density for the alternating orbital ordering calculated by DMFT (QMC). The parameters are $U'=3.5\,$eV, $T=0.05\,$eV and $g=0.05\,$eV$^{3/2}$.}

Fig.\ \ref{Fig:sec:undoped:DMFT:polarAB} plots the orbital polarization 
$P=|\sum_{\sigma}(n^A_{1\sigma}-n^A_{2\sigma})|$, i.e., the difference of orbital occupations $n^A_{\mu\sigma}$, as a function of bandwidth and temperature. For $W=3.6\,$eV, we find a finite orbital polarization below $T\approx 725\,$K, in agreement with the experimental result of the structural transition temperature $T_{OO}=740\,$K \cite{Chatterji2003}. 
\tig[clip=true,width=5.5cm,angle=270]{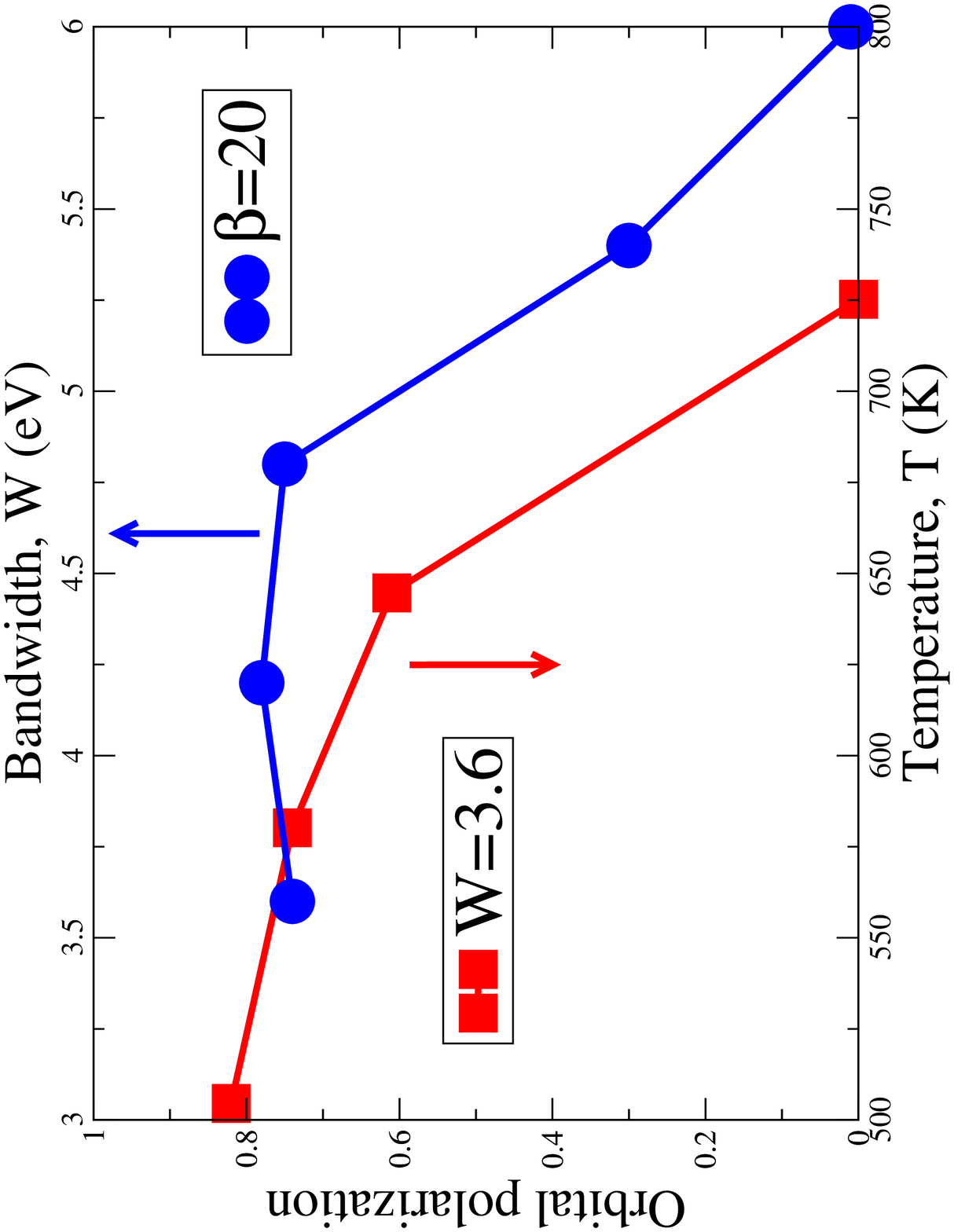}
{Fig:sec:undoped:DMFT:polarAB}
{Orbital polarization vs.\ temperature at fixed $W=3.6\,$eV (squares) and vs.\ bandwidth at fixed $T=0.05\,$eV (circles). The Jahn-Teller coupling is $g=0.05\,$eV$^{3/2}$ and the inter-orbital Coulomb repulsion $U'=3.5\,$eV.} 
If we fix the temperature $T=0.05\,$eV and var the bandwidth as under pressure, the orbital polarization is almost a constant for the bandwidth $W<4.8\,$eV and then decreases with increasing bandwidth until it is reduced to zero at $W=6.0\,$eV. This behavior reflects the nature of the bandwidth-control metal-insulator transition in LaMnO$_3$. Here, $W=4.8\,$eV corresponds to the critical bandwidth where the split minority and majority $e_g$ bands start to overlap; and LaMnO$_3$ becomes metallic. Hence, for a larger bandwidth, the orbital polarization decreases rapidly with increasing bandwidth.

The bandwidth $W=6.0\,$eV marks a second transition where the orbital
polarization and the Jahn-Teller distortion are completely suppressed. Although different from Ref.\ \onlinecite{Loa2001}, we can still identify three distinct regimes at low temperature: (i) an insulating phase with orbital ordering and static Jahn-Teller distortion below $W\approx 4.8\,$eV (or $P_{IM}=32\,$GPa); (ii) an intermediate metallic phase with orbital ordering and static Jahn-Teller distortion below $W\approx 6\,$eV; (iii) a metallic phase with orbital symmetry and dynamic Jahn-Teller distortion above $W\approx 6\,$eV.

\section{Doped managanites}
\label{Sec:doped}
\subsection{Paramagnetic insulating state}
\label{Sec:doped:PI}

Let us now turn to the doped manganites with their extraorinary properites such as the colossal magnetoresistance. For simplicity, we first neglect the breathing phonon mode. We start by plotting the probability distribution of
the lattice distortion for $n=0.8$ ($x=0.2$) at different temperatures and
couplings in Fig.\ \ref{Fig:sec:doped:dmft:polaron}. The Coulomb interaction is fixed to $U'=3.5\,$eV throught Section \ref{Sec:doped}. In contrast to the single peak distribution obtained for the undoped case in 
Fig.\ \ref{Fig:sec:undoped:DMFT:distribution}, we find two peaks in the distribution function at large Jahn-Teller couplings for doped systems. The large peak located at about $g/\Omega^2$ corresponds to the large lattice distortion due to the Jahn-Teller coupling (and corresponds to the single peak without doping), while the small peak stems from the quantum and thermal fluctuation of the MnO$_6$ octahedra which due to the doping are not occupied with an electron (and hence not strongly split by the electron-phonon coupling $g$). While at a large coupling strength $g\gtrsim 0.1$, the two peaks are well separated, they merge into a single structure at weak coupling such as for $g=0.05\,$eV$^{3/2}$ in Fig.\ \ref{Fig:sec:undoped:DMFT:distribution}.

\tig[clip=true,width=8cm,angle=270]{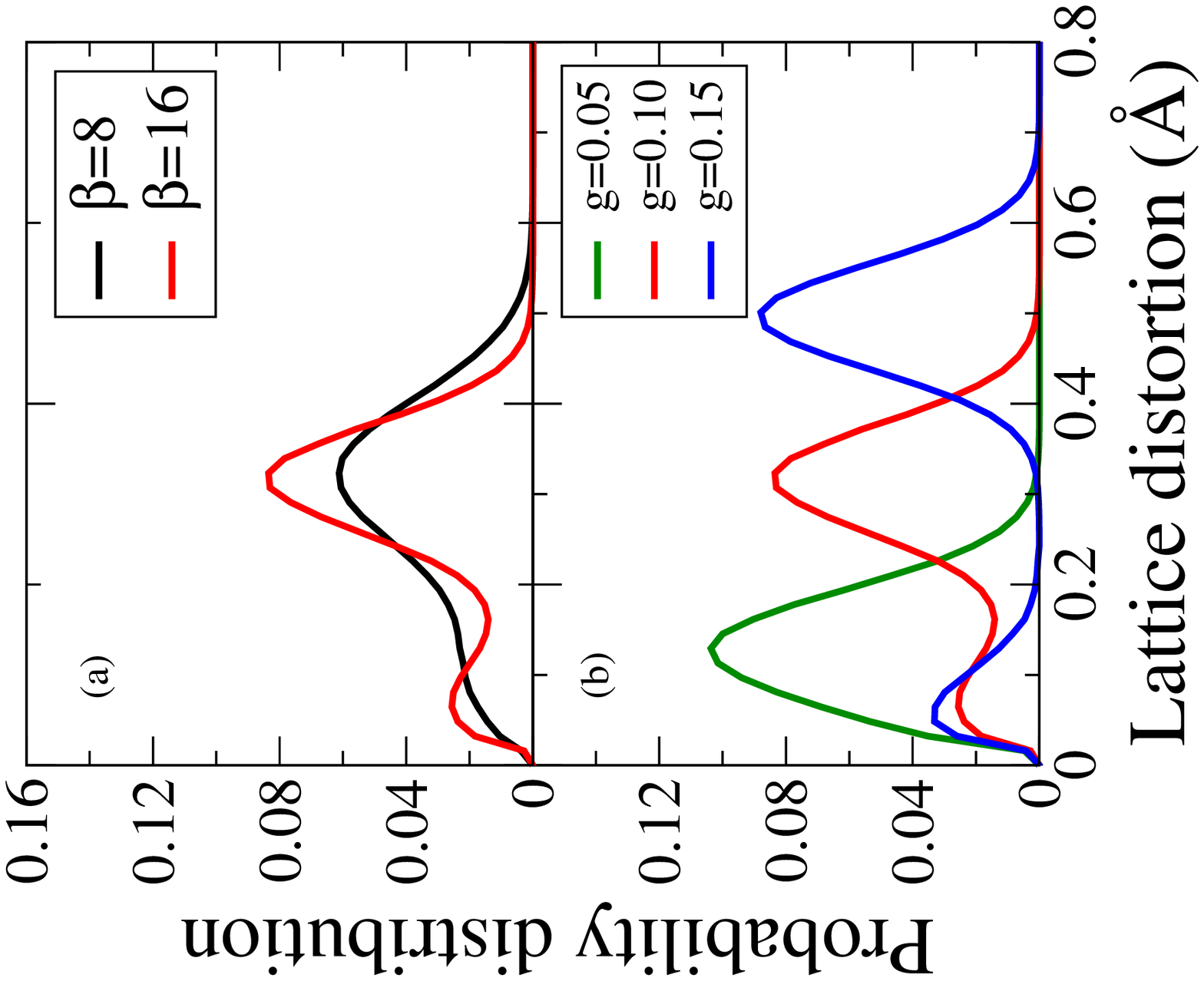}
{Fig:sec:doped:dmft:polaron}
{Probability distribution of the lattice distortion at $n=0.8$ $e_g$ electrons per site (i.e., doping $x=0.2$) for (a) $g=0.10\,$eV$^{3/2}$ and $\beta=8$ and $16\,$eV$^{-1}$ and (b) $\beta=16\,$eV$^{-1}$ and $g=0.05$, $0.10$, and $0.15\,$eV$^{3/2}$.
}

The connection to the electronic spectrum is shown in 
Fig.\ \ref{Fig:sec:doped:dmft:polaron2}. For clarity, we separate the spectrum into two parts originated from the large and small lattice distortions, respectively. For the large lattice distortion, the corresponding spectral density locates well below or far above the Fermi energy. The low energy part can be identified as localized $e_g$ electrons or more precisely the transition from this state to states without $e_g$ electron. The high energy part stems from the states which are pushed up by the Hund's rule coupling, the Jahn-Teller splitting and the Coulomb repulsion. For the small lattice distortion, most of the spectral density locates slightly above the Fermi energy. They stem from the "undistorted" unoccupied states which are also called "midgap states" in the literature \cite{Millis1996a,Jung1998}.

The formation of lattice polarons provides for the basic physics of doped
manganites. The electron spectral density can be seen as a combination of
polaron states well below the Fermi energy and the midgap states above the Fermi energy. As shown in Fig.\ \ref{Fig:sec:doped:dmft:polaron2}, this results in the strong suppression of the spectral weight at the Fermi energy and gives rise to a large energy gap for $g=0.15\,$eV$^{3/2}$ and a pseudo-gap for smaller Jahn-Teller couplings.

\tig[clip=true,width=7cm,angle=270]{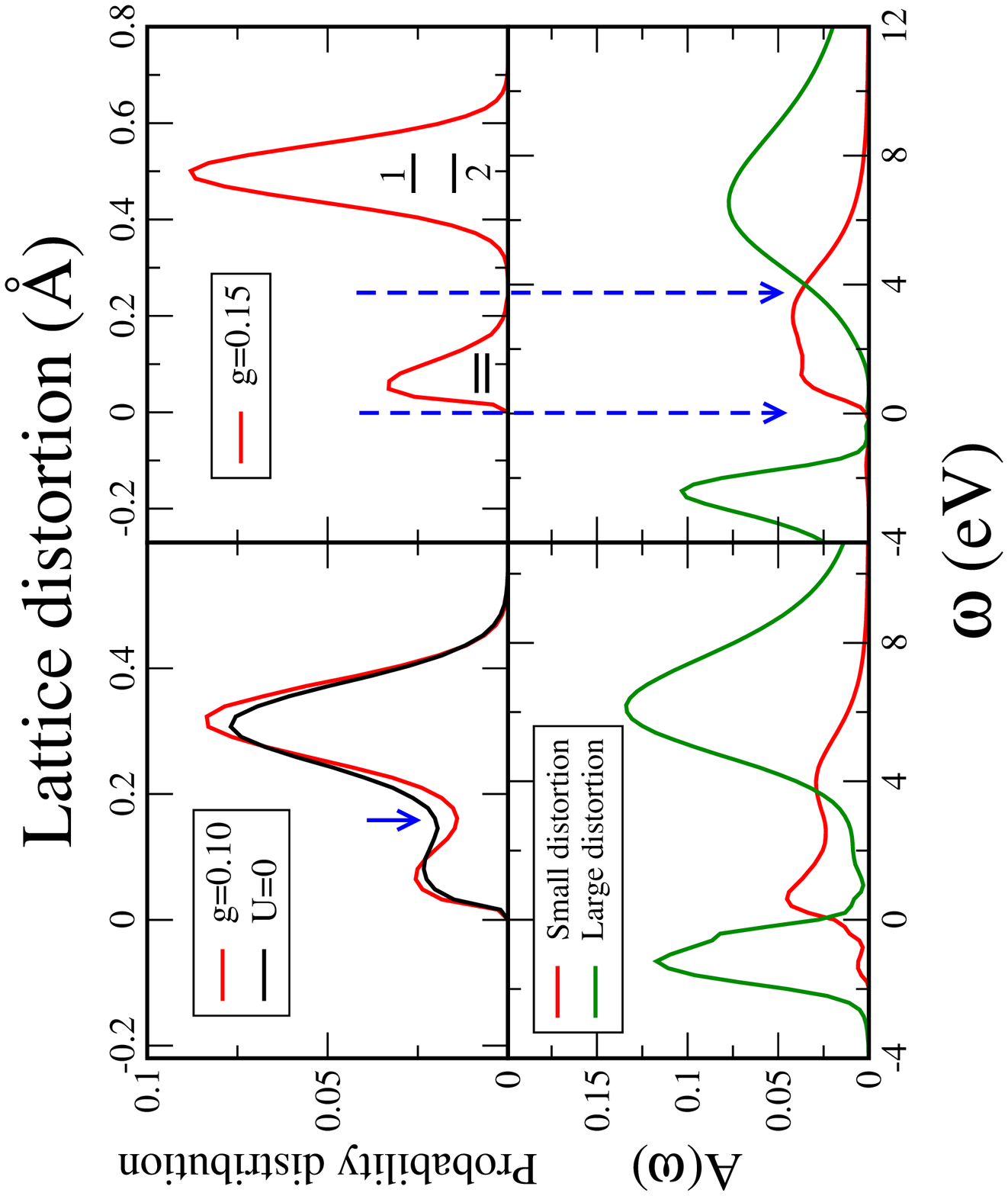}
{Fig:sec:doped:dmft:polaron2}
{Probability distribution of the lattice distortion and the corresponding
electron spectral densities for the Jahn-Teller coupling $g=0.10\,$eV$^{3/2}$ (left panel) and $0.15$ (right panel) at inverse temperature $\beta=16\,$eV$^{-1}$ and $n=0.8$ $e_g$ electrons per site. The electron spectral densities are separated into two parts with large and small lattice distortions. The arrows indicate the separation between peaks. The phonon distribution at $g=0.10\,$eV$^{3/2}$ without the Coulomb interaction is also shown for comparison.
}

Let us discuss the role of the on-site Coulomb interaction in the formation
of the pseudo gap and hence the insulating-like paramagnetic state of manganites. Fig.\ \ref{Fig:sec:doped:dmft:compareU} compares the spectral densities for $g=0.10\,$eV$^{3/2}$ with and without the Coulomb interaction; Fig.\ \ref{Fig:sec:doped:dmft:polaron2} the corresponding probability
distribution of the lattice distortion. For $U=5\,$eV ($U'=3.5\,$eV), some spectral weight is pushed away from the Fermi energy to higher energies and the pseudo-gap is strongly enhanced by the Coulomb interaction. Also the lattice distortion shows a more pronounced separation of the two peaks which indicates the enhancement of the polaron formation by the Coulomb interaction. This confirms the important role of the Coulomb interaction.

\tig[clip=true,width=6cm,angle=270]{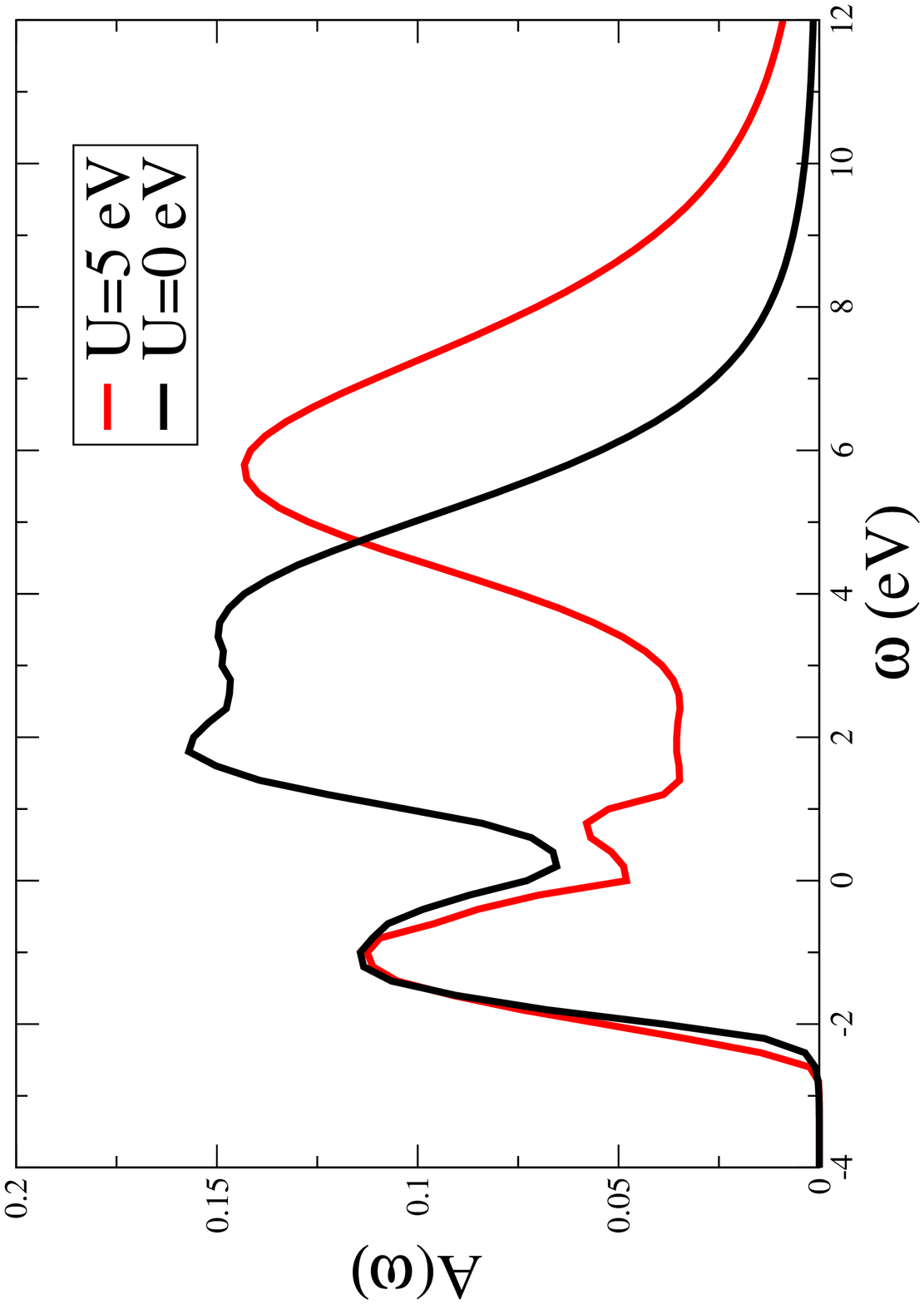}
{Fig:sec:doped:dmft:compareU}
{Paramagnetic electronic spectral density at $n=$0.8, $\beta=16\,$eV$^{-1}$ and $g=0.10\,$eV$^{3/2}$ comparing $U=0\,$eV and $U=5\,$eV ($U'=3.5\,$eV).
}

Let us now consider the gradual swiching on of the Jahn-Teller couplings $g$  at fixed $\beta=16\,$eV$^{-1}$ and $n=0.8$ shown in 
Fig.\ \ref{Fig:sec:doped:dmft:spectrum}. Without Jahn-Teller coupling, the spectrum has a broad quasiparticle peak at the Fermi energy and the system is metallic. With increasing $g$, the quasiparticle peak is gradually suppressed and the spectral weight at the Fermi energy is reduced, leaving a dip in the spectrum for intermediate coupling and an energy gap for strong coupling. The low (high) energy Hubbard band also shifts towards lower (higher) energies. An additional peak shows up at $\omega=1-2\,$eV, ascribed to the midgap states with small dynamic lattice distortion.

\tig[clip=true,width=6cm,angle=270]{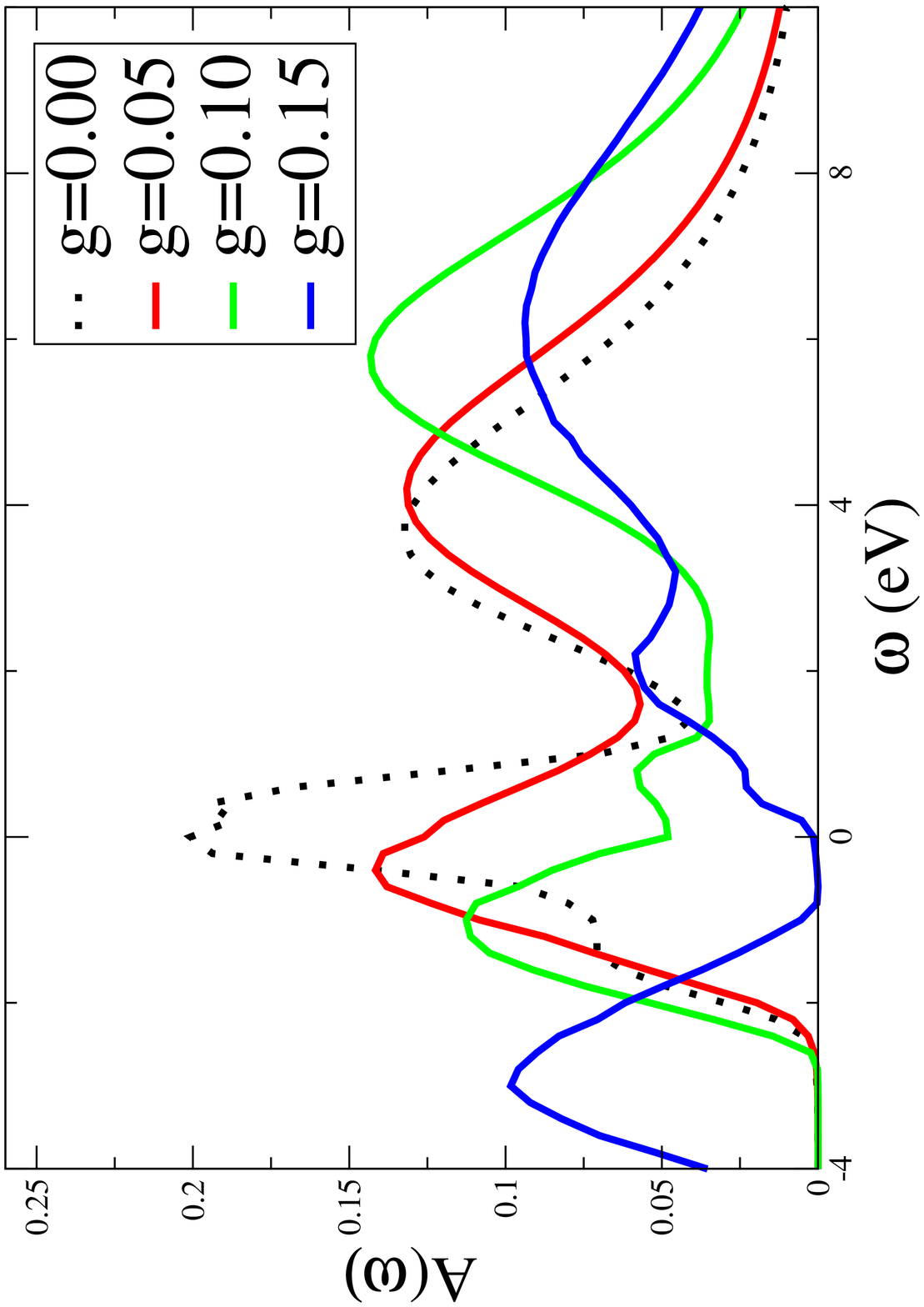}
{Fig:sec:doped:dmft:spectrum}
{Electron spectral density in the paramagnetic phase for different couplings $g=0$, 0.05, 0.10, and 0.15 eV$^{3/2}$. The parameters are $\beta=16\,$eV$^{-1}$ and $n=0.8$. The quasiparticle peak is suppressed at large $g$, giving the
pseudo-gap behavior for intermediate coupling and a large energy gap for strong coupling.
}

Turning to the doping dependence at fixed $\beta=16\,$eV$^{-1}$ and $g=0.10\,$eV$^{3/2}$ in Fig.\  \ref{Fig:sec:doped:dmft:spectrumN}, we see that the system is a good insulator with a large energy gap of about 2.3 eV without doping. This is larger than the experimental gap of 1 eV and hence suggests that
a smaller  Jahn-Teller coupling such as $g=0.05\,$eV$^{3/2}$ used in Section \ref{Sec:Parent} is more realistic. But for the qualitative discussion here, we continue with $g=0.1\,$eV$^{3/2}$ since for this larger Jahn-Teller coupling
the different features of the spectrum can be better identified. A small doping $x=0.1$ introduces some midgap states just above the Fermi energy and changes the large energy gap into a pseudo-gap at the Fermi energy. If the doping is large enough, the midgap states can dominate at the Fermi energy so that the pseudo-gap is completely filled at $n=0.3$. However, due to the strong phonon and spin scattering, the quasiparticle peak is still damped (broadened); i.e., the life time of the quasiparticles is very short.

\tig[clip=true,width=6cm,angle=270]{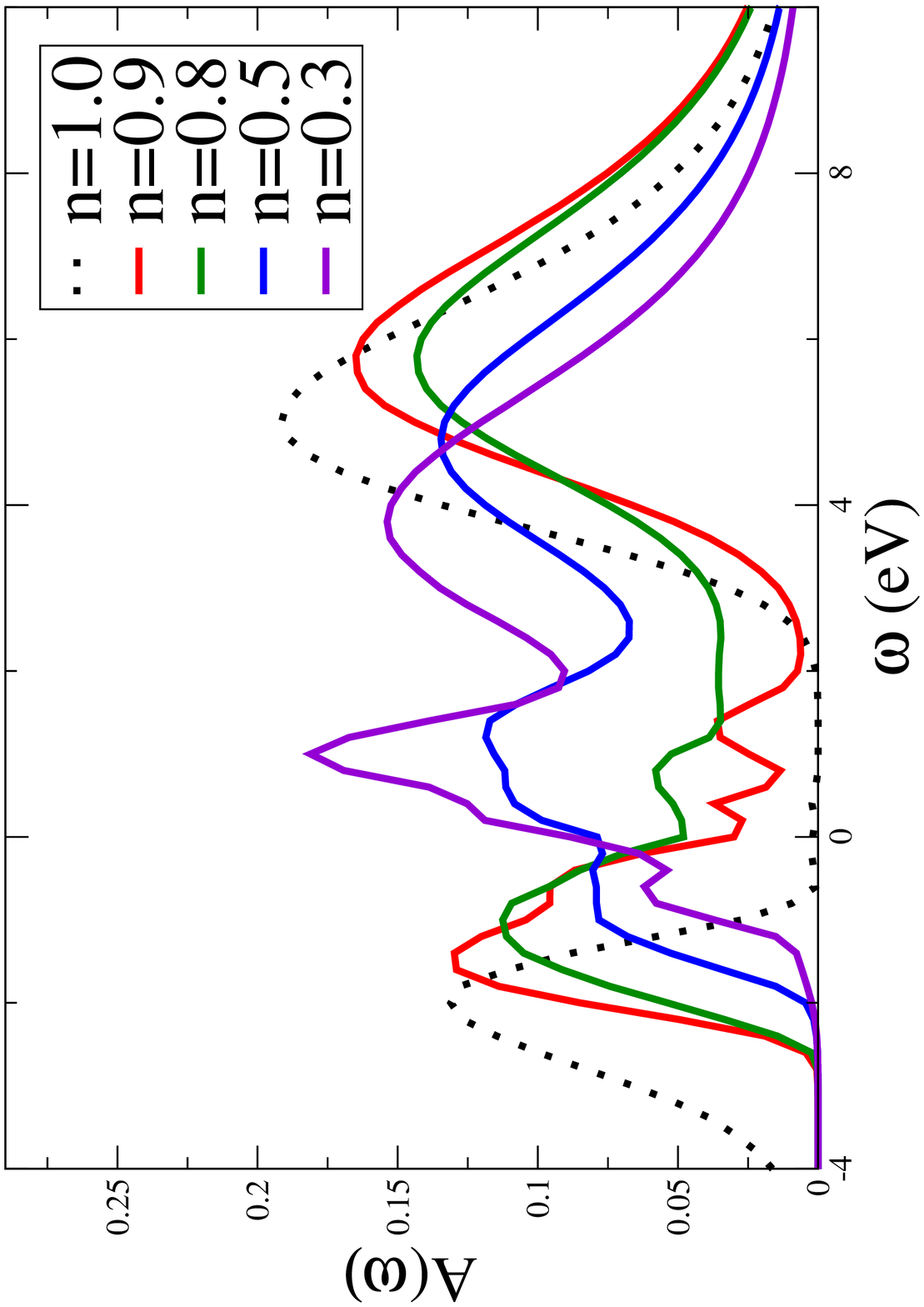}
{Fig:sec:doped:dmft:spectrumN}
{Paramagnetic spectral density for different electron occupations $n=$1.0, 0.9, 0.8, 0.5, and 0.3 at $\beta=16\,$eV$^{-1}$ and $g=0.10\,$eV$^{3/2}$.
}

Since the temperature dependence of the spectrum was already discussed in 
Ref.\ \onlinecite{Yang2006}, we restrict ourselves here to the optical conductivity in Fig.\ \ref{Fig:sec:doped:dmft:opcondthT}. 
\fig[clip=true,width=5.5cm,angle=270]{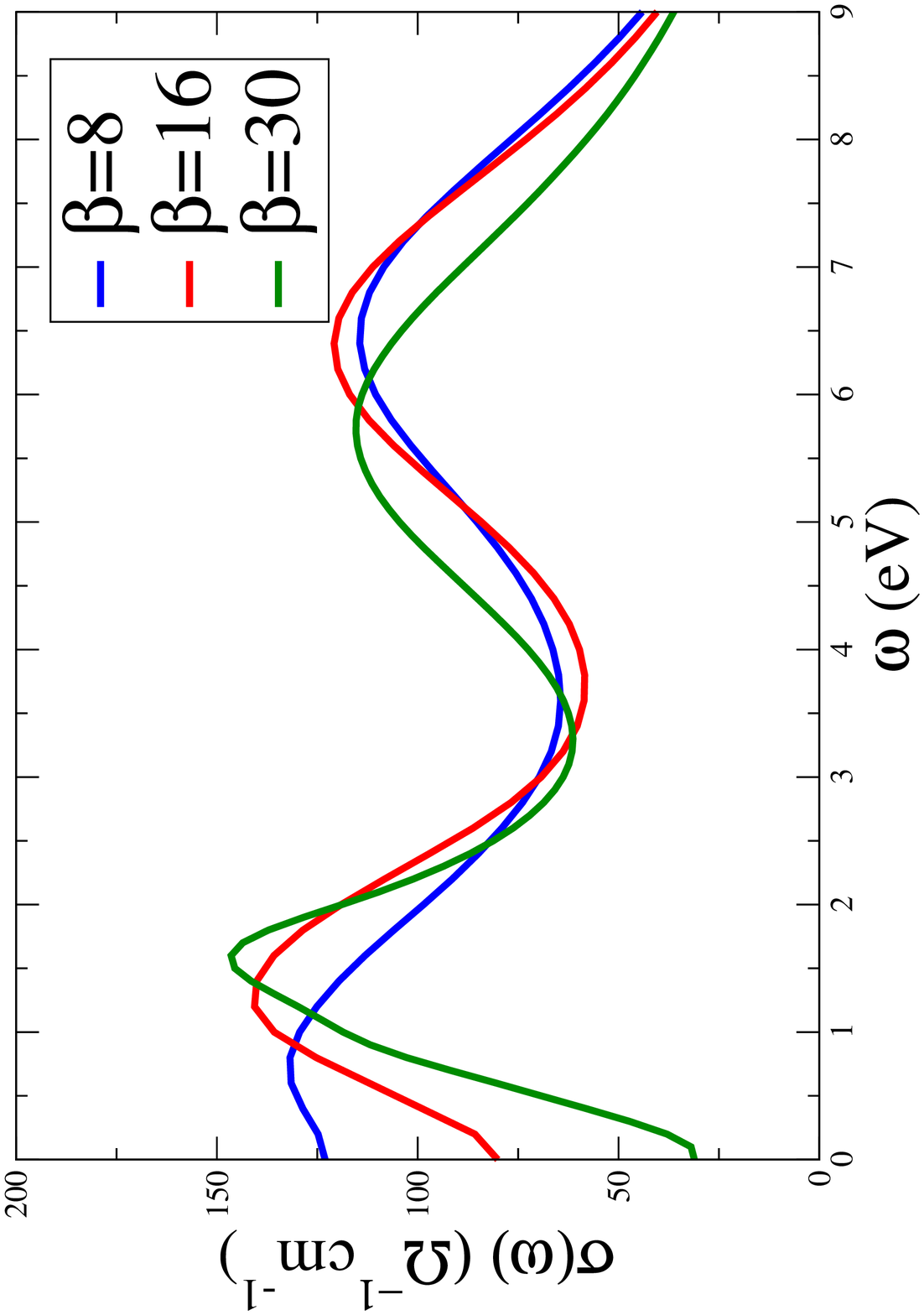}
{Fig:sec:doped:dmft:opcondthT}
{Optical conductivity in the paramagnetic phase at average occupancy $n=0.8$ and inverse temperature $\beta=8$, 16, and 30 eV$^{-1}$.
}
The low energy optical conductivity is suppressed as a result of the pseudo-gap in the spectral density. With decreasing temperature, the optical spectrum is more suppressed since there are less thermal excitations and the two peak lattice distortion becomes more pronounced 
(Fig.\ \ref{Fig:sec:doped:dmft:polaron}) as does the pseudo gap \cite{Yang2006}. Hence, the resistivity is  enhanced at low temperature, giving rise to the insulating behavior in the paramagnetic phase, see Ref.\ \onlinecite{Yang2006} for the $T$-dependence of the resistivity.

The two peaks of the optical spectrum stem from the two peaks in the electronic spectrum above ther Fermi energy. An excitation of an electron into the unoccupied, undistorted  states just above the Fermi energy gives rise to the midgap states around $1.5\,$eV. Transitions to the distorted states far above the Fermi energy with two $e_g$ electrons are at the origin of the peak around 6 eV.

\subsection{Ferromagentic phase transition}
\label{Sec:doped:FM}
The low temperature ferromagnetic phase is a bad metal which due to the strong spin and phonon scattering has quasiparticles with very short life times and no true Drude peak \cite{Yang2006}. Here, we will concentrate on the ferromagnetic phase transition which can be either driven by applying a magnetic field or by decreasing the temperature. Fig.\ \ref{Fig:sec:doped:dmft:Tc} shows the doping dependence of the Curie temperature for the Jahn-Teller coupling
$g=0.10\,$eV$^{3/2}$. Compared to the experimental results, the theoretical predictions of $T_c$ have a similar shape with a maximum at intermediate doping but overall the values of $T_c$ are about 2-3 times larger in magnitude. Close to the undoped parent compound ($x=0$, $n=1$), the Curie temperature is suppressed since the Coulob interaction hinders the movement of the electrons
so that the double exchange is no longer effective; instead antiferromagnetism prevails \cite{Held2000}.

\tig[clip=true,width=6cm,angle=270]{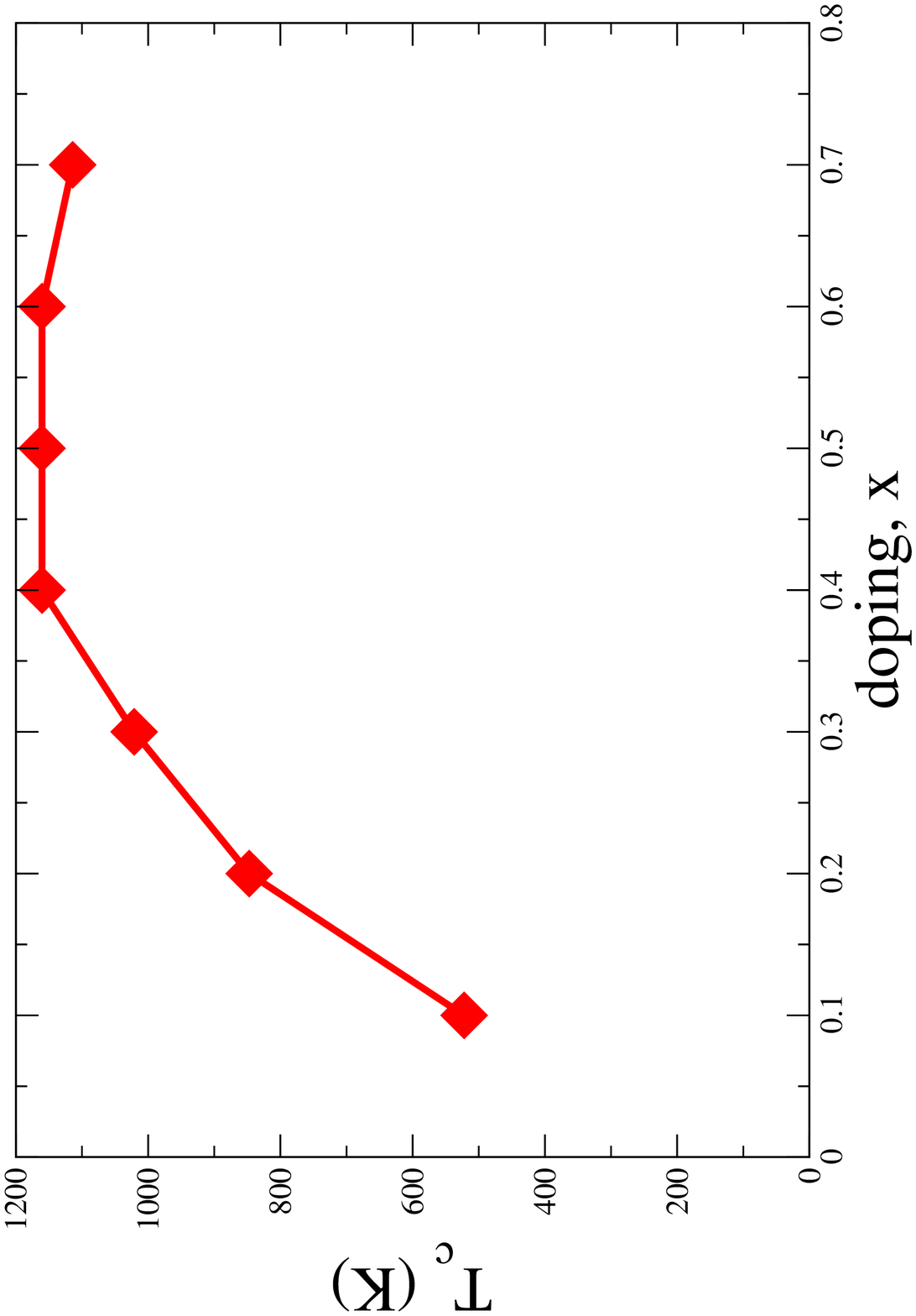}
{Fig:sec:doped:dmft:Tc}
{Curie temperature as a function of doping $x=1-n$ for $g=0.10\,$eV$^{3/2}$. 
}

The magnitude of the Cure temperature depends strongly on the strength of the
Jahn-Teller coupling and the hopping integral of the $e_g$ electrons. 
Fig.\ \ref{Fig:sec:doped:dmft:Tcg} plots its coupling dependence at $n=0.8$, which are similar to the previous results \cite{Millis1996c,Edwards2002}. The Curie temperature decreases rapidly with increasing coupling. For strong Jahn-Teller coupling, a slight increase of $g$ from 0.10 to 0.12 (or the dimensionless coupling $\lambda$ from 2.0 to 2.2) reduces $T_c$ by a factor of 2 so that $g=0.12\,$eV$^{3/2}$ (or $\lambda=2.2$) gives the correct experimental value.

\tig[clip=true,width=6cm,angle=270]{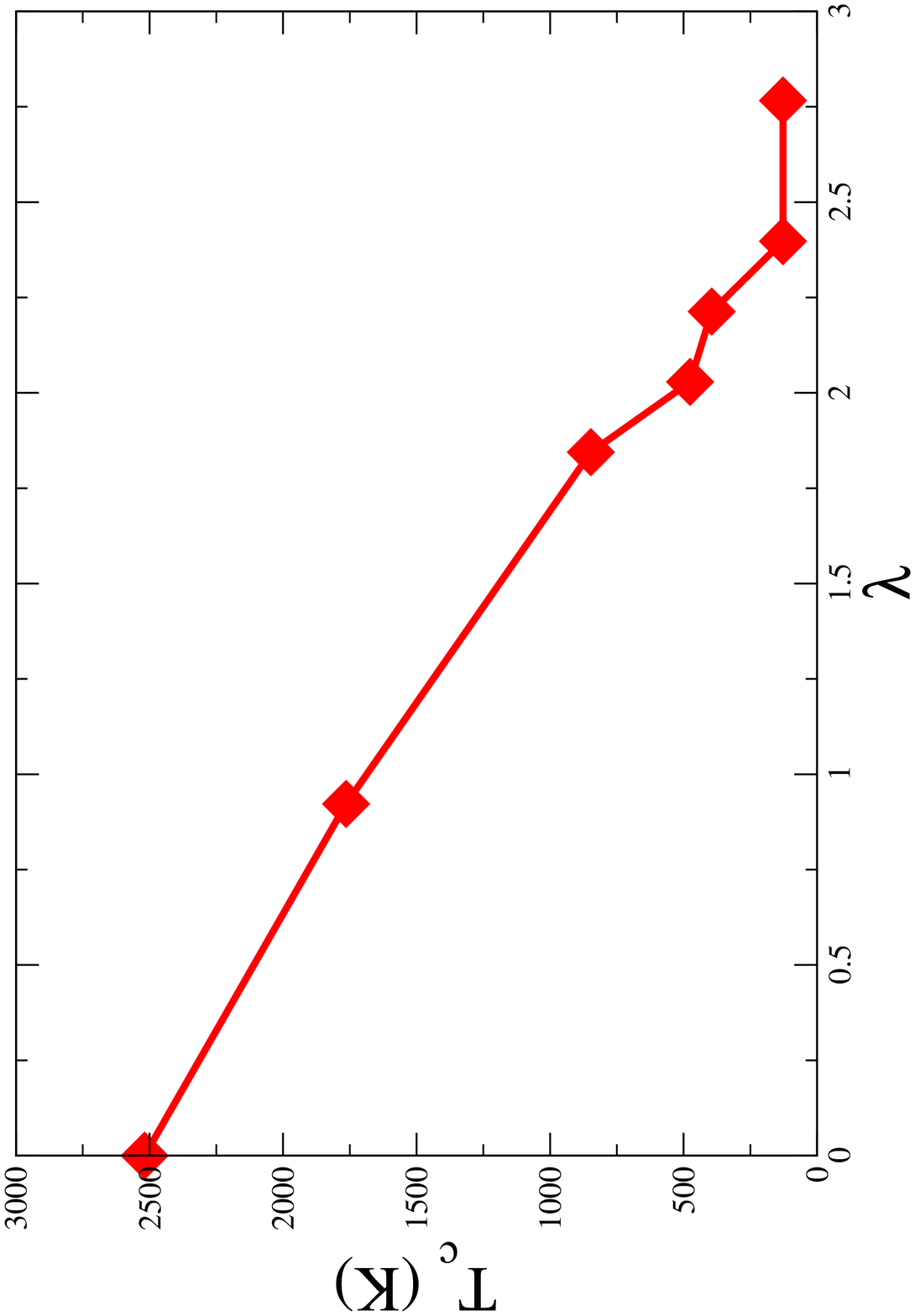}
{Fig:sec:doped:dmft:Tcg}
{Curie temperature as a function of the dimensionless coupling $\lambda=g/\Omega\sqrt{t_0}$ at $n=0.8$.
}

The Curie temperature may also be affected by many factors which are not included in our low-energy Hamiltonian. These include the quantum fluctuation of Mn $t_{2g}$ spins and the antiferromagnetic superexchange coupling between Mn $t_{2g}$ spins. Both tend to weaken the ferromagnetic order and suppress the Curie temperature. The antiferromagnetic superexchange coupling has been estimated to be the order of 200 K \cite{Perring1997,Dagotto2001} which, if taken into account, would greatly reduce the theoretical value of the Curie temperature. Also the mean field character of the DMFT approximation tends to overestimate the Curie temperature.

In the literature, the Curie temperature has been calculated with different methods such as DMFT \cite{Millis1995,Millis1996a,Held2000}, conventional mean-field theory \cite{Roder1996}, QMC simulations \cite{Yunoki1998}, and the many-body CPA \cite{Edwards2002}. Some of the results seem to be in better agreement with experiments. However, we should note that these results are all based on the details of the models and approaches and are very sensitive to the values of the parameters which, unfortunately, are not always reliable and, as a matter of fact, vary considerably in the literature. A complete analysis of
the problem is still required.

\section{Breathing mode}
\label{Sec:Breathing}
In this section we include a third electron-phonon coupling, the breathing mode which couples to the electron density, see section \ref{Sec:Model}.
\gig[clip=true,width=5.5cm,angle=270]{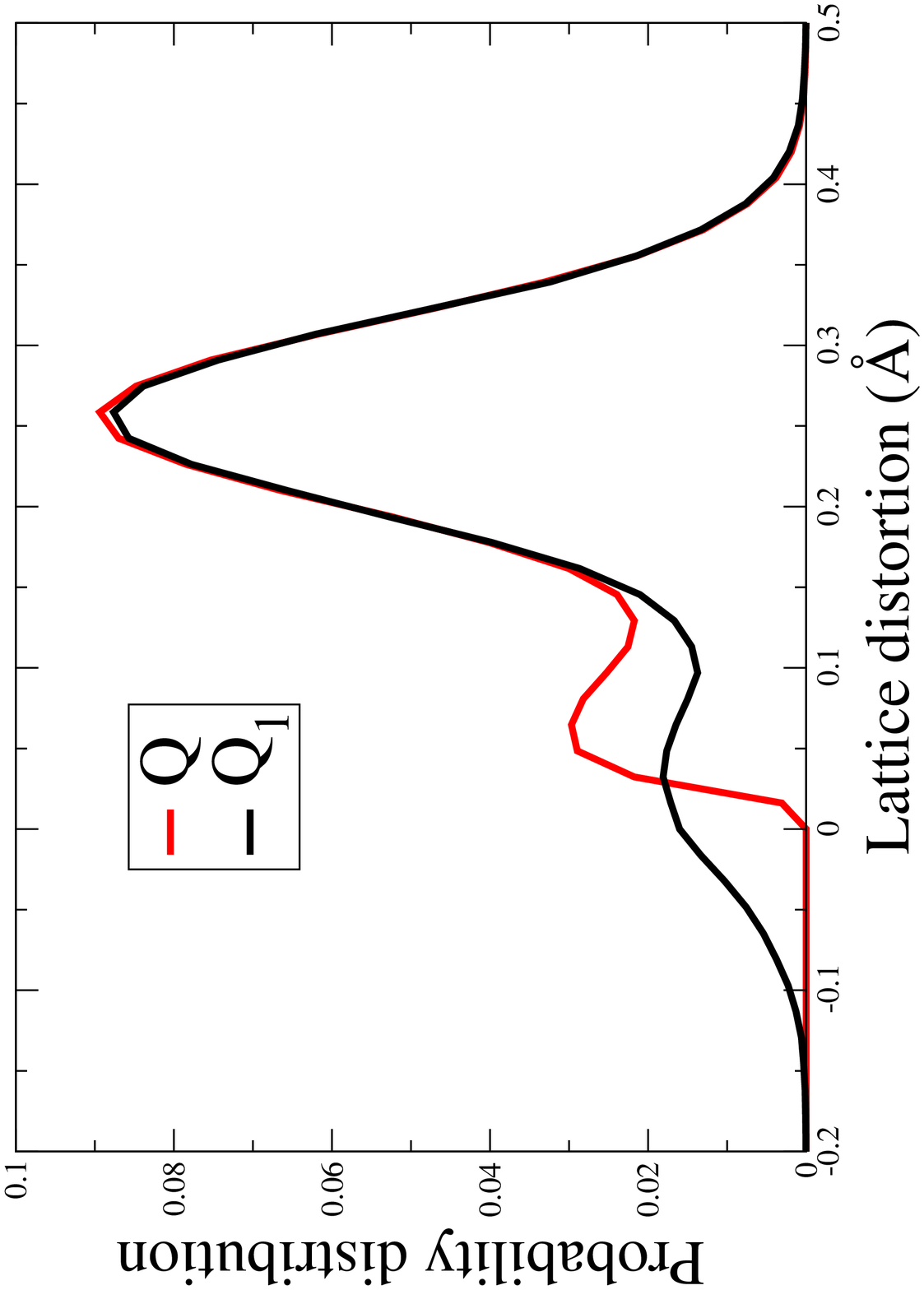}
{Fig:sec:doped:dmft:phonon3mode}
{Probability distribution of the breathing ($Q_1$) and Jahn-Teller ($Q$) modes for the model Hamiltonian (\ref{Eq:sec:doped:dmft:parameters:Ham}). The
parameters are $n=0.8$ and $\beta=16\,$eV$^{-1}$, $U=5\,$eV, $U'=3.5\,$eV, $J=0.75\,$eV, $g=0.08\,$eV$^{3/2}$, and $\Omega=0.07\,$eV for all three modes.
}
Fig.\ \ref{Fig:sec:doped:dmft:phonon3mode} shows the probability distribution of both the breathing and Jahn-Teller phonon fields at $\beta=16\,$eV$^{-1}$, $n=0.8$,  $g=0.08\,$eV$^{3/2}$, $U=5\,$eV, $J=0.75\,$eV and $U'=3.5\,$eV, and a phonon frequency $\Omega=0.07\,$eV for all three modes. Except for the Jahn-Teller distortion $Q$ being positive by definition, the distribution function of the breathing phonon is similar to that of the Jahn-Teller distortion and  also has two peaks located at about $g/\Omega^2$ and $0\,$. These two peaks are related to the polaron states and the midgap states discussed before. The only difference is the way in which the phonon modes are coupled to the $e_g$ electrons: The breathing mode is coupled to the electron density, while the Jahn-Teller modes are coupled to the orbital polarization (with respect to a certain basis). Due to the Hund's coupling and the strong Coulomb interaction, double occupations are forbidden. Hence, the difference is not reflected in the distribution function. 

Since the breathing mode only couples to the electron density, it lowers
the localization energy of the polaron states but leaves the midgap states
unchanged. The $e_g$ electrons are thus more localized due to the inclusion of
the breathing mode and the system becomes more insulating.

Fig.\ \ref{Fig:sec:doped:dmft:modescomp} compares the spectral densities at
$\beta=16\,$eV$^{-1}$ and $n=0.8$ with and without the breathing phonon. As expected, we see the density of states at the Fermi energy to be strongly suppressed by the existence of the breathing mode, demonstrating how the breathing mode supports the tendencies towards localization of the $e_g$ electrons.

\gig[clip=true,width=5.5cm,angle=270]{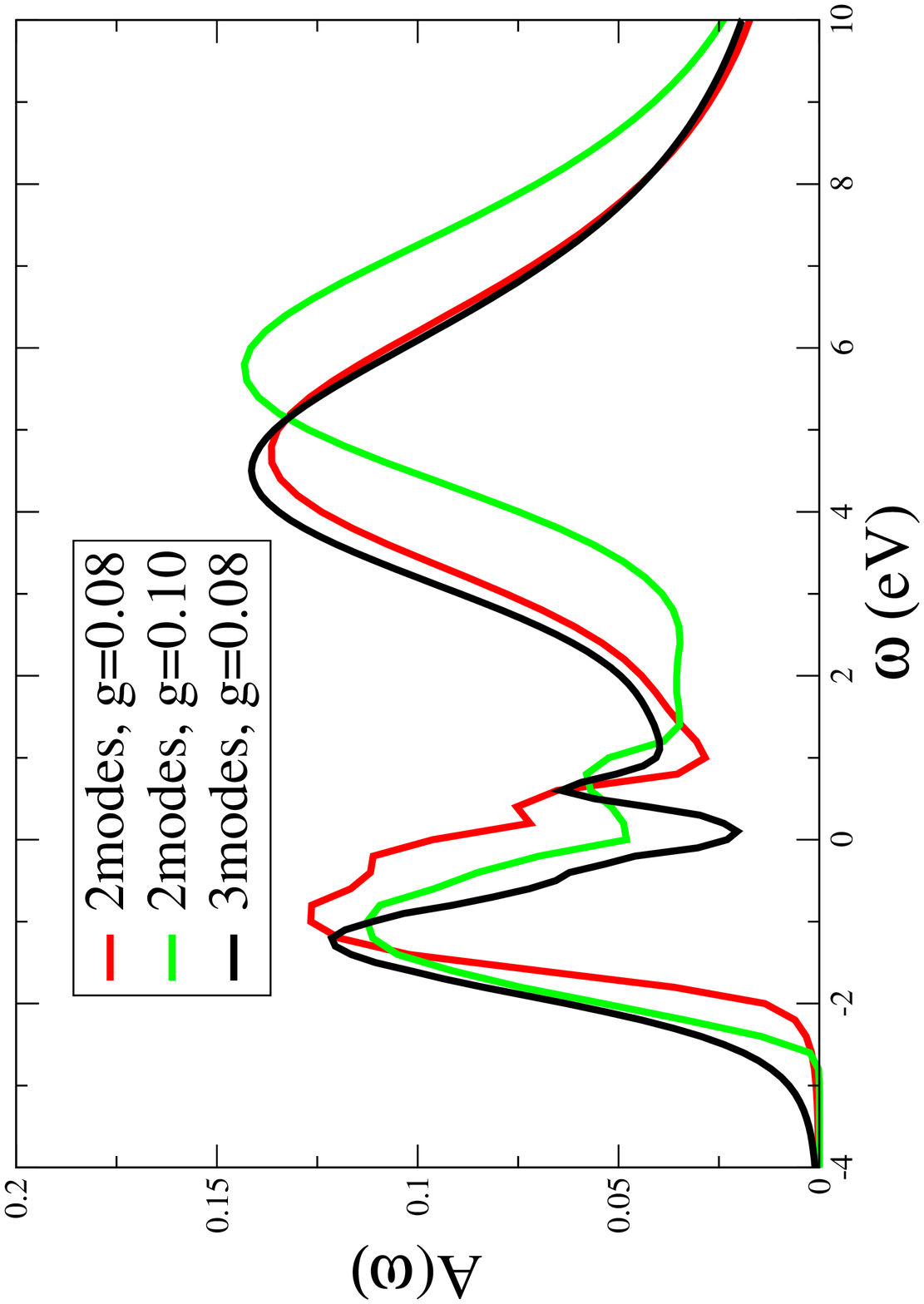}
{Fig:sec:doped:dmft:modescomp}
{Electron spectral density for $U'=3.5\,$eV, $n=0.8$ and $\beta=16\,$eV$^{-1}$. The coupling constant is taken as $g=0.08\,$eV$^{3/2}$ for calculations with all three phonon modes and $g=0.08$, $0.10\,$eV$^{3/2}$ for calculations with only the Jahn-Teller modes. The phonon frequencies are taken as $\Omega=0.07\,$eV for all three modes.
}

\subsection{Determing $U$ and $g$}
\label{Sec:Breathing:Parameters}
Let us now try to estimate from our results including the breathing mode the two parameters with the hitherto largest uncertaity, i.e., the on-site Coulomb interaction $U$ and the electron-phonon coupling $g$. All other parameters are fxed as discussed in Section \ref{Sec:Model}. To this end, we fit the two parameters to two experimental values, i.e., the experimental gap of undoped LaMnO$_3$ which is 1 eV and the resitivity for $x=17.5\%$ Sr  doping which is  $0.035\,\Omega{}$cm. The calculations are done for  $\beta=30\,$eV$^{-1}$.

\gig[clip=true,width=6.cm,angle=270]{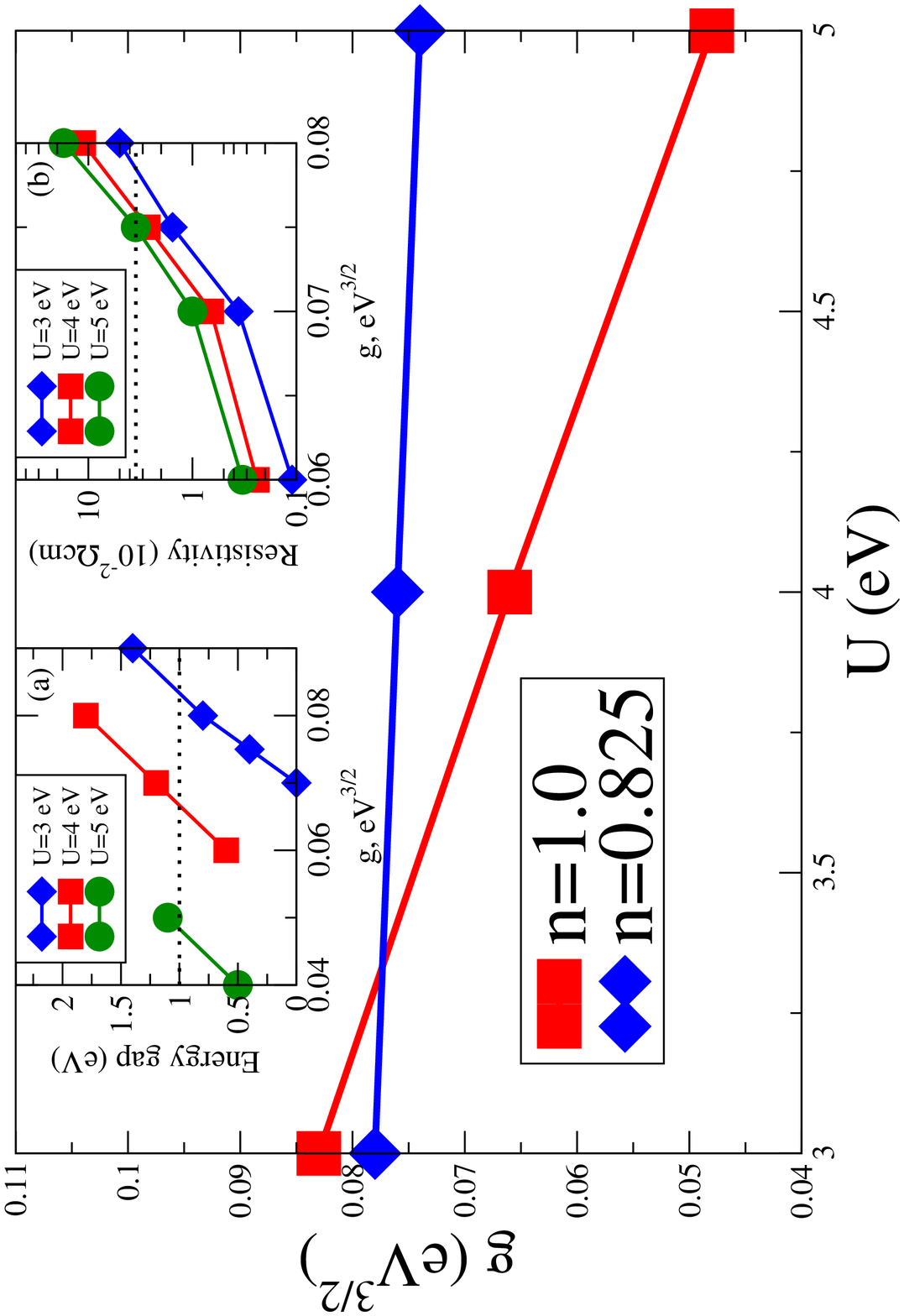}
{Fig:sec:doped:dmft:ug}
{Two sets of parameters which fit the experimental energy gap in LaMnO$_3$ and the resistivity in La$_{0.825}$Sr$_{0.175}$MnO$_3$. Their intersection gives a
single set ($U$,$g$) which is expected to describe quantitatively both doped and undoped manganites. The insets show the DMFT (QMC) results as a function of the Coulomb interaction $U$ and the Jahn-Teller coupling $g$ for: (a) the energy gap at $n=1$ and (b) the resistivity at $n=0.825$. Both are calculated at $\beta=30\,$eV$^{-1}$. The experimental results are indicated by the dotted line with the 1 eV energy gap in LaMnO$_3$ and the resistivity of about $0.035\,\Omega{}$cm for La$_{0.825}$Sr$_{0.175}$MnO$_3$ \cite{Tokura1994,Urushibara1995}.
}

The inset (a) of Fig.\ \ref{Fig:sec:doped:dmft:ug} plots the energy gap for different Coulomb interactions $U$ and Jahn-Teller couplings $g$ calculated by DMFT (QMC) for the model Hamiltonian (\ref{Eq:sec:doped:dmft:parameters:Ham}). As expected, the energy gap depends strongly on both parameters.

The inset (b) of Fig.\ \ref{Fig:sec:doped:dmft:ug} shows the DMFT (QMC) results of the resistivity for different $U$ and $g$ at $n=0.825$ ($x=0.175$) and
$\beta=30\,$eV$^{-1}$. As has been analyzed before, the Coulomb interaction also
affects the resistivity of the system since it enhances the electron 
localization. This is now proved explicitly in the figure. The resistivity
increases with increasing $U$, albeit it depends more sensitive on $g$ than on $U$.

By comparing with the experimental data (dotted lines in the two insets), 
the two experiments provide for two different sets of $U$ and $g$ parameters which we plot in the main panel of Fig.\ \ref{Fig:sec:doped:dmft:ug}. Only the intersection at $U=3.3\,$eV and $g=0.077\,$eV$^{3/2}$ agrees with both experiments. These parameters determined by us agree with the crude estimates of 
Section \ref{Sec:Model}. We hence expect this paremeter set, summarized in Table  \ref{Tab:sec:doped:parameters}, to be the proper set for the model Hamiltonian (\ref{Eq:sec:doped:dmft:parameters:Ham}), describing both doped and undoped
manganites.

\begin{table}[t]
\begin{center}
\begin{tabular}{|c|c|c|c|c|c|}
\hline
\rule[-3mm]{0mm}{8mm} W & U & $2{\cal J}|{\bf S}|$ & J & $\Omega$ & g \\
\hline
\rule[-3mm]{0mm}{8mm} $3.6\,$eV  &  $3.3\,$eV  & $2.7\,$eV & $0.75\,$eV  &  $0.07\,$eV  &
$0.077\,$eV$^{3/2}$ \\
\hline
\end{tabular}\\
\end{center}
\caption{Parameters estimated for doped and undopd manganites. $W=6t_0$: bandwidth; $U$: intra-orbital Coulomb interaction; $J$: $e_g$-$e_g$ Hund's exchange; ${\cal J}$: $e_g$-$t_{2g}$ Hund's coupling; $\Omega$: phonon frequency; $g$: Jahn-Teller coupling. The bandwidth is obtained from the LDA calculations for the cubic structure \cite{Yamasaki2006}, the Hund's coupling is calculated by the constrained LDA for the ferromagnetic phase, and the phonon frequency is estimated from the Raman spectroscopy \cite{Iliev1998}. Only the Coulomb interaction $U$ and the Jahn-Teller coupling $g$ are estimated from the DMFT (QMC) calculations for the model Hamiltonian (\ref{Eq:sec:doped:dmft:parameters:Ham}) by fitting the experimental data of the energy gap in LaMnO$_3$ and the resistivity in La$_{0.825}$Sr$_{0.175}$MnO$_3$, see Fig.\ \ref{Fig:sec:doped:dmft:ug}.}
\label{Tab:sec:doped:parameters}
\end{table}

\tig[clip=true,width=5.5cm,angle=270]{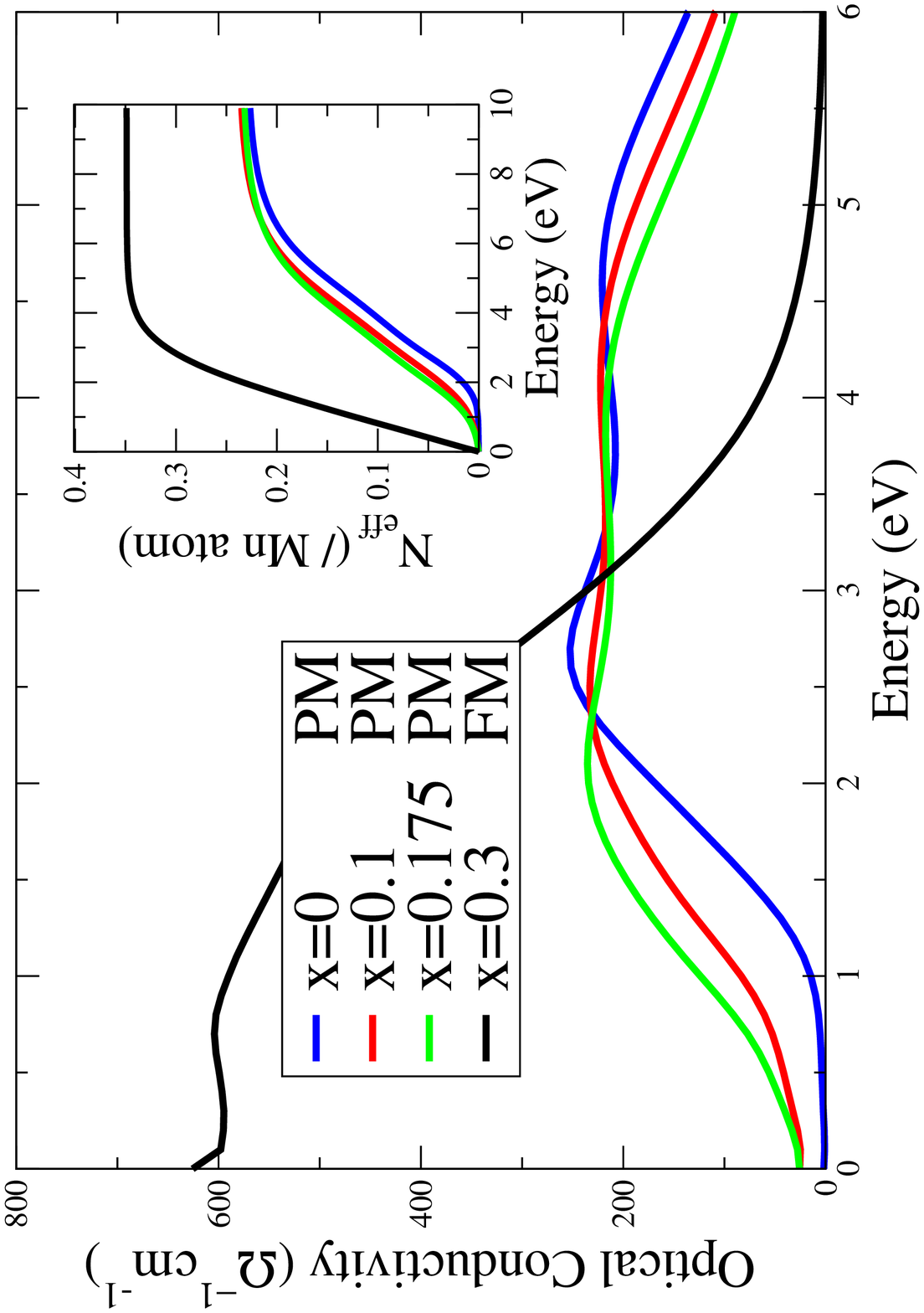}
{Fig:sec:doped:dmft:optug}
{Optical conductivity at doping $x=0$, $0.1$, $0.175$ in the paramagnetic (PM) phase and at $x=0.3$ in the ferromagnetic (FM) phase. The parameters are taken as is in Table \ref{Tab:sec:doped:parameters} and the inverse temperature is $\beta=30\,$eV$^{-1}$. The inset shows the effective carrier concentration calculated from the integrated optical conductivity.
}

\subsection{Optical conductivity}
\label{Sec:Breathing:Optics}
The optical conductivity with the breathing mode included for the determined set of paramaters is presented in Fig.\ \ref{Fig:sec:doped:dmft:optug}. The inset is the effective carrier concentration $N_{eff}(\omega)$ calculated from the integral of the optical conductivity which can be compared directly to the experimental results of  Ref.\ \onlinecite{Takenaka1999}. We see a continuous crossover from the undoped to the doped system and a good agreement in the general behavior of the optical conductivity with experiment. However, the
experimental magnitude of the optical peaks are 2-3 times larger than the
theoretical one \cite{Quijada1998}. This gives rise to an unexpected large experimental $N_{eff}(\omega)$ even below 4 eV, in contrast to the theoretical predictions presented in the inset of Fig.\ \ref{Fig:sec:doped:dmft:optug}. This suggests that, besides the $e_g$ states also oxygen $p$ and Mn $t_{2g}$ states
which we did not account for in the theoretical calculation contribute to the experimental optical conductivity. Another possible explanation is the insufficiency of the group velocity as a substitute for the more apropriate dipole matrix elements.

\section{Conclusion}
\label{Sec:Conclusion}
We have used DMFT to study the physics of manganites by hands of  a realistic microscopic model which takes into account both the electron-electron and electron-phonon interactions, together with the Hund's rule coupling between the $e_g$ conduction electrons and the $t_{2g}$ spins. In the undoped system, the model produces similar results as the previous LDA+DMFT calculations and, most surprisingly, it also predicts the correct structural transition temperature  
from dynamic to static Jahn-Teller distortion. In the doped phase, the $e_g$ electrons are trapped by the large lattice distortion and form a lattice polarons. This process is strongly supported by the Coulomb interaction. Our results provide for an explanation of the insulating-like paramagnetic state  over a wide range of doping. The CMR is a result of a transition towards a ferromagnetic (bad) metallic state at a Curie temperature which shifts by applying an external manetic field. The dynamical properties are determined by the polaron states and midgap states which stem from the undistorted unoccupied
states slightly above the Fermi eenrgy. The combination of both gives rise to the pseudo-gap behavior observed in doped manganites. The inclusion of the breathing mode further favors the  tendencies towards polaron formation.
Our results show that the realistic microscopic model can be applied to both doped and undoped manganites and can therefore be taken as the starting point towards a complete understanding of the physics of manganites in which the cooperative effect of the Jahn-Teller phonons and the quantum fluctuation and superexchange coupling of the $t_{2g}$ spins must also be taken into account for quantitative comparison with experiment. For future studies we provide for a realistic set of model parameters in Table \ref{Tab:sec:doped:parameters}.

\section*{Acknowledgements}
\label{Sec:Ack}
This work has been supported by the EU-Indian cooperative FP-7 network MONAMI;
work at Los Alamos was performed under the auspices of the US Department of Energy.

\end{document}